\documentclass[fleqn,usenatbib]{mnras}

\usepackage{newtxtext,newtxmath}

\usepackage[T1]{fontenc}

\DeclareRobustCommand{\VAN}[3]{#2}
\let\VANthebibliography\thebibliography
\def\thebibliography{\DeclareRobustCommand{\VAN}[3]{##3}\VANthebibliography}

\usepackage{graphicx}	
\usepackage{amsmath}	
\usepackage{threeparttable} 
\usepackage{subcaption}
\usepackage{url}
\usepackage{hyperref}
\usepackage{ulem}

\title[Comet 3I/ATLAS]{Post-perihelion Photometric and Spectroscopic Monitoring of the Interstellar Comet 3I/ATLAS}

\author[Goldy Ahuja et al.]{
Goldy Ahuja$^{1,2}$\thanks{E-mail: goldy@prl.res.in | goldezz20@gmail.com},
Shashikiran Ganesh$^{1}$,
K. Aravind$^{3}$,
Emmanuel Jehin$^{3}$,
Bhavya Ailawadhi$^{1}$,
\newauthor
Vikrant K. Agnihotri$^{4}$,
B. Arvind$^{1,2}$,
Pallavi Saraf$^{1}$,
Alka Singh$^{1}$,
Deekshya R. Sarkar$^{1}$,
Sunil Chandra$^{1}$,
\newauthor
Prithish Halder$^{5}$,
Devendra Sahu$^{6}$,
T. Sivarani$^{6}$,
and Anil Bhardwaj$^{1}$
\\
$^{1}$Physical Research Laboratory, Ahmedabad, Gujarat-380009, India\\
$^{2}$Indian Institute of Technology Gandhinagar, Palaj, Gujarat-382355, India\\
$^{3}$Space sciences, Technologies \& Astrophysics Research (STAR) Institute, University of Liège, B-4000 Liège, Belgium\\
$^{4}$Cepheid Observatory, Rawatbhata, Rajasthan-323307, India\\
$^{5}$Department of Physics and Astronomy, University of Nebraska-Lincoln, Lincoln, NE, 68588-0299, USA\\
$^{6}$Indian Institute of Astrophysics, Koramangala, Bengaluru, Karnataka – 560034, India
}

\date{Accepted XXX. Received YYY; in original form ZZZ}

\pubyear{\the\year{}}

\begin{document}
\label{firstpage}
\pagerange{\pageref{firstpage}--\pageref{lastpage}}
\maketitle

\begin{abstract}
The third interstellar object, comet 3I/ATLAS, was observed using various Indian telescopes during its post-perihelion evolution. We carried out imaging and spectroscopic observations of the comet at multiple epochs using the 0.5--m, 1.2--m, and 2.5--m telescopes of the Physical Research Laboratory and the 2-m Himalayan Chandra Telescope (HCT) operated by the Indian Institute of Astrophysics. Additionally, time-series photometry was obtained on 7 epochs using the 0.132--m refractor of Cepheid Observatory. The spectroscopic observations of 3I/ATLAS reveal signatures of molecular emission from CN, C$_2$, and C$_3$, as commonly observed in the emission spectra of solar system comets. The production rates and rate ratios indicate that the composition of the 3I/ATLAS transitioned from carbon-depleted (pre-perihelion) to carbon-typical (post-perihelion). The change in carbon composition could be due to exposure of fresh material from subsurface layers near perihelion. We also performed photometric observations to determine broadband colours and the spin period. The measured colours are $B-V=$~1.03~$\pm$~0.13 mag, $V-R=$~0.59~$\pm$~0.11 mag, $R-I=$ 0.08~$\pm$~0.04 mag, $g-r=$~0.52~$\pm$~0.09 mag, and $r-i=$~0.13~$\pm$~0.08 mag. The colours are redder than the Long-Period Comets of the solar system and the Sun in $B-V$ and $V-R$, but near solar in $R-I$. The measured colours remain nearly constant throughout the observing period and are consistent with the reported pre-perihelion values. The colours and the reflectivity gradient match the photometric colours of the `ultrared-material rich' objects in the solar system. The nuclear radius derived from non-gravitational force lies between 0.45 and 0.65 km.
\end{abstract}




\section{Introduction}
Comets are relatively small icy bodies composed of volatile-rich materials and dust, typically ranging in size from sub-kilometres to a few kilometres \citep{Meech_2004}. They are considered remnants of the early solar system that were dynamically scattered during the migration of the giant planets \citep{Gomes2005,Tsiganis2005,Morbidelli_2007,Walsh2011,Brasser_2013}. As a result, these primitive bodies were redistributed into distant reservoirs, primarily the Kuiper Belt and the Oort Cloud. The Kuiper Belt, extending beyond Neptune at distances of $\sim$50--100 au, is the main source region of Short-Period Comets (SPCs), most of which move in low-inclination orbits close to the ecliptic plane. In contrast, the Oort Cloud is a distant spherical reservoir extending from $\sim$2000 to 230000 au \citep{Oort_1950,Chebotarev_1965}, and is believed to be the source of Long-Period Comets (hereafter LPCs) and exhibits nearly isotropic orbital inclinations. Since the LPCs spend most of their lifetime at large heliocentric distances, their subsurface layers are expected to preserve relatively pristine material from the early solar system. As these bodies approach the Sun, sublimation of volatile species produces a gaseous coma and dust tail, allowing their chemical composition to be studied through spectroscopic observations.

In a similar way, there are other minor bodies that form in distant stellar systems and might be ejected from their parent systems \citep{Jewitt_ANRev}. To date, three interstellar objects (hereafter ISO) have been observed in our solar system. The first, 1I/$'$Oumuamua (C/2017 U1), an asteroid-like object with no detectable coma \citep{Meech_1I_2017,Quan_1I_2017,Oumuamua_ISSI_2019}. The second ISO, 2I/Borisov (C/2019 Q4), was well studied due to its visible coma, and its properties are similar to solar system comets with a high content of CO, which were well explained by many authors \citep{Opitom_2I_2019,Fitzsimmons_2I_2019, Jewitt_2I_2020, Piotr_2I_Nature2020,Bodewits_2I_2020,Aravind_2I_2021,Ahmed_2021, prodan_2I_2024}. In July 2025, the third ISO, 3I/ATLAS (C/2025 N1), hereafter 3I, was discovered at a heliocentric distance of $4.51$ au, with an eccentricity of $6.13 \pm 0.02$ \citep{3I_disc_1,3I_disc_2}. The comet was moving with a very high hyperbolic excess velocity of around $58$ km s$^{-1}$ and reached perihelion at $1.357 \pm 0.001$ au on 2025 October 29. In comparison, the two other ISOs had hyperbolic excess velocities of $26.33$ km s$^{-1}$ and $32.3$ km s$^{-1}$, respectively, making comet 3I the fastest ISO identified to date. Several kinematic studies have been conducted to study the origin of this comet. \citet{Taylor_2025} suggests that the comet may be part of the thick disk, with the parent star to be metal-poor. The study by \citet{Fuente_3I} favour that it originated in the thin disk, while \citet{Hopkins_2025} and \citet{Ahuja_2025} found that the comet is in the transition region between the thin and thick disks.

At the time of discovery, the spectrum of comet 3I was devoid of gaseous emission, and its reflectance spectrum was significantly redder than those of the other two known ISOs. The spectral gradient was estimated to be $\sim$16--22\% per 1000 \AA\ over the 4000--7000 \AA\ wavelength range similar to that of D-type asteroids \citep{3I_disc_1,3I_disc_2,Opitom_3I_disc,Belyakov_3I_disc,Fuente_3I,Tony_2025,Hoogendam_2025c}. \citet{Salazar_Manzano_2025} estimated a slope of $26.1 \pm 0.5$\% per 1000 \AA\ in 3900-6000 \AA\}.\citet{Yang_2025} measured it to be $14.1 \pm 1.3$\% per 1000 \AA\ and $11.4 \pm 0.2$\% per 1000 \AA\, which is lower than the values found by different authors. Using photometric observations, several authors calculated the pre-perihelion colours of comet 3I \citep{3I_disc_1,Opitom_3I_disc,Alvarez_2025} and reported a very red colour, with $B-V$ ranging from 0.98 mag to 1.12 mag and $V-R$ ranging from 0.51 mag to 0.57 mag. However, \citet{3I_disc_1} reported a comparatively redder value of $V-R = 0.71~\pm~0.09$ mag. Similarly, \citet{Opitom_3I_disc} measured a $R-I$ of $0.53~\pm~0.17$ mag, whereas \citet{3I_disc_1} found a nearly neutral colour of $0.14~\pm~0.10$ mag. In addition, from photometric light curves, \citet{Fuente_3I} estimated the nucleus rotation period of the comet 3I to be $16.79 \pm 0.23$ h, and \citet{Tony_2025} estimated the nucleus rotation period of the comet 3I to be $16.16 \pm 0.01$ h.

From the spectroscopic point of view, \citet{Alvarez_2025} obtained the spectrum using the X-Shooter instrument on 2025 July 04 at a distance of 4.4 au and did not detect a clear signature of CN emission. However, they provided an upper limit of the production rate of CN (Q$_{\rm CN}$) to be $5.6$ $\times$ 10$^{23} ~ \rm s^{-1}$. Subsequently, \citet{Salazar_Manzano_2025} observed comet 3I at heliocentric distances of 3.2–-2.9 au and first detected the clear CN emission with a production rate (Q$_{\rm CN}$) of $\sim 6 \times 10^{24} ~ \rm s^{-1}$. They further measured the upper limit of log$_{10}$(C$_2$/CN) to be less than $-$0.8, indicating that the chemical composition of comet 3I was highly carbon-depleted. This could be due to inefficient sublimation of C$_2$ at the observed heliocentric distance. Similarly, \citet{Hoogendam_2025c}, using the Keck Cosmic Web Imager mounted on the Keck-II at a heliocentric distance of 2.75 au on 2025 Aug 24, detected CN emission with a high signal-to-noise ratio. However, the C$_2$ emissions remained undetected.  Around the same period (2025 August 11--21), from the X-shooter instrument, \citet{Rahatgaonkar_2025} measured the CN production rate in the range $(1.28-6.24) \times 10^{24}$ s$^{-1}$. Additional pre-perihelion observations at a heliocentric distance of 3.14 au to 1.85 au by \citet{Hutsemekers_2026} with the UVES and X-SHOOTER instruments further confirmed that the comet was carbon-depleted.

In November, comet 3I became observable again during its post-perihelion phase, and  \citet{Ganesh_17502_ATel_2025} reported the first post-perihelion spectroscopic observations. Subsequently, \citet{Hoogendam_2026} obtained the optical spectra of the comet and detected multiple emission bands, including CN, C$_2$, C$_3$, as well as Ni\,I and Fe\,I lines. The derived log$_{10}$(C$_2$/CN) ratio of $-0.26~\pm~0.14$ showed a significant change compared to its pre-perihelion values, but still remained between the typical and carbon-depleted classes defined by \citet{AHEARN_1995}. In contrast, the measured log$_{10}$(C$_3$/CN) ratio of $-1.51~\pm~0.14$ placed the comet within the carbon-chain depleted region. Using low-resolution spectroscopic observations obtained between December 2025 and January 2026 at a heliocentric distance of 1-8--3.3 au with multiple instruments, \citet{Zhao_2026} reported the median post-perihelion value of log$_{10}$(C$_2$/CN) to be $-0.10~\pm~0.05$ and log$_{10}$(C$_3$/CN) of $-1.06~\pm~0.09$, suggesting that the chemical composition of the comet 3I had changed from highly carbon depleted to a more carbon-typical composition. Similarly, multi-epoch observations by \citet{Kawakita_2026}, at heliocentric distances of 1.77 to 1.95 au, indicated that the cometary composition evolved from highly depleted to moderately depleted, with log$_{10}$(C$_2$/CN) varying between $-0.24$ to $-0.40$. Also, \citet{Jehin_17515_2025,Jehin_17538_2025} observed the comet at heliocentric distances of 1.76 and 1.98 au using the Hale-Bopp (HB) narrowband photometric filters on TRAPPIST telescopes and measured the log$_{10}$(C$_2$/CN) ratio to be 0.04. These post-perihelion observations confirm the change in the composition.

In this paper, we present the results from the photometric and spectroscopic observations of comet 3I obtained during its post-perihelion journey using multiple Indian observatories. In section \ref{sec: observations}, we will discuss the observation details from different instruments using different telescopes, while section \ref{sec: data reduction} outlines the methodology and data reduction procedures. In section \ref{sec: Results}, we will present the results of these observations and compare them with the pre-perihelion values, followed by the conclusions in section \ref{sec: Conclusion}. As the comet 3I goes away from the Sun, it offers a unique opportunity to investigate the post-perihelion evolution of the composition and activity of this interstellar comet.

\begin{table*}
\caption{Log of spectroscopic observations obtained with different Indian telescopes.}
\setlength{\tabcolsep}{4.5pt}
\begin{threeparttable}
\resizebox{\textwidth}{!}{%
\begin{tabular}{lcccccccccc}
\hline
\multicolumn{1}{c}{{Date}} &
\multicolumn{1}{c}{{Instrument/}}&
\multicolumn{1}{c}{{TFP\tnote{a}}} &
  \multicolumn{1}{c}{{r$_h$\tnote{b}}}&
  \multicolumn{1}{c}{{$\Delta$\tnote{c}}} &
  \multicolumn{1}{c|}{{Time}}&
  \multicolumn{1}{c}{{Observation}}&
  \multicolumn{1}{c}{{Exposure}}&
  \multicolumn{1}{c}{{Phase}}&
  \multicolumn{1}{c}{{Airmass}} &
  \multicolumn{1}{c}{{Standard}}\\
\multicolumn{1}{c}{} &
  \multicolumn{1}{c}{{Telescope}} &
\multicolumn{1}{c}{(days)} &
  \multicolumn{1}{c}{{(au)}} &
  \multicolumn{1}{c}{{(au)}} &
  \multicolumn{1}{c}{{(UT)}} &
  \multicolumn{1}{c}{{Type}} &
  \multicolumn{1}{c}{{Time (s)}} &
  \multicolumn{1}{c}{{Angle ($^{\circ}$)}} &
  \multicolumn{1}{c}{{}}&
  \multicolumn{1}{c}{{Star}}\\\hline
2025 Nov 16 & LISA / 1.2$-$m PRL & 16.5 & 1.50 & 2.09 & 23:48 & Spectroscopy & 900 & 26.27 & 2.65 & Feige 66\\
2025 Nov 17 & LISA / 1.2$-$m PRL & 17.5 & 1.50 & 2.09 & 23:51 & Spectroscopy & 1200 & 27.92 & 2.86 & Feige 66\\
2025 Nov 27 & HFOSC / 2.0$-$m HCT & 28.5 & 1.75 & 1.95 & 20:36 & Spectroscopy (Gr7) & 900 & 30.25 & 1.62 & BD+284211\\
2025 Nov 28 & HFOSC / 2.0$-$m HCT & 29.5 & 1.75 & 1.95 & 23:28  & Spectroscopy (Gr7) & 1200 & 30.51 & 1.60 & HILT600\\
2025 Dec 04 & LISA / 1.2$-$m PRL & 35.5 & 1.87 & 1.88 & 23:42 & Spectroscopy & 1200 & 30.42 & 1.32 & Feige 56\\
2025 Dec 09 & LISA / 1.2$-$m PRL & 40.5 & 2.00 & 1.84 & 23:20 & Spectroscopy & 1200 & 29.31 & 1.23 & Feige 56\\
2025 Dec 12 & HFOSC / 2.0$-$m HCT & 43.5 & 2.08 & 1.82 & 23:41 & Spectroscopy (Gr7) & 1200 & 28.19 & 1.49 & BD+284211\\
2025 Dec 13 & LISA / 1.2$-$m PRL & 44.5 & 2.11 & 1.81 & 20:47 & Spectroscopy & 1200 & 27.75 & 2.10 & Feige 56\\
2025 Dec 17 & LISA / 1.2$-$m PRL & 48.5 & 2.22 & 1.80 & 20:58 & Spectroscopy & 1200 & 25.66 & 1.66 & Feige 56\\
2025 Dec 20 & HFOSC / 2.0$-$m HCT & 51.5 & 2.31 & 1.80 & 20:50 & Spectroscopy (Gr7) & 1800 & 23.81 & 1.69 & BD+284211\\
2026 Jan 20 & HFOSC / 2.0$-$m HCT & 83.5 & 3.29 & 2.31 & 16:45 & Spectroscopy (Gr7) & 1800 & 1.25 & 1.23 & BD+75d325\\
\hline
\end{tabular}
}
\begin{tablenotes}
\footnotesize
\item $^{\rm a}$ Time from Perihelion. $^{\rm b}$ Heliocentric Distance. $^{\rm c}$ Geocentric Distance.
\end{tablenotes}
\end{threeparttable}
\label{tab: observation_details}
\end{table*}

\section{Observations}\label{sec: observations}
We have observed comet 3I using various telescopes, i.e., the $0.5-$m, $1.2-$m and $2.5-$m Physical Research Laboratory (hereafter PRL\footnote{\url{https://www.prl.res.in/~miro/telescopes.html}}) Mount Abu telescopes, the $2~$m Himalayan Chandra Telescope (hereafter HCT\footnote{\url{https://www.iiap.res.in/centers/iao/facilities/hct/}}), and the 0.132$-$m telescope from the Cepheid Observatory (hereafter CepO\footnote{\url{https://posts.3cepheids.co.in/wp/}}). The spectroscopic and photometric observations were carried out from the HCT and PRL telescopes, while only photometric observations were conducted from the CepO. Detailed descriptions of the HCT and PRL telescopes are given in \citet{Ahuja_2025_MNRAS} and \citet{Kumar_2014}. In the next subsections, we will briefly discuss the details of the instruments and telescopes used in this work.

\subsection{0.5-m, 1.2-m and 2.5-m telescopes at PRL Mount Abu Observatory}\label{PRL}
We conducted spectroscopic and photometric observations using multiple instruments at the Mount Abu Observatory (Lat.: 24.65$^\circ$ N; Long.: 72.78$^\circ$ E; Alt.: 1680 m). The spectroscopic observations were obtained using the Long-slit Intermediate Resolution Spectrograph for Astronomy (hereafter LISA) mounted on the 1.2$-$m PRL $f/13$ telescope, while the photometric observations were carried out using the Wide-Field Imager (hereafter WFI), a CMOS-based imager installed on both the 0.5$-$m $f/6.8$ and 1.2$-$m PRL $f/13$ telescopes. Additional photometric observations were carried out using the Faint Object Camera (hereafter FOC\footnote{\url{https://www.prl.res.in/~miro/instruments.html}}) mounted on the 2.5$-$m PRL $f/8$ telescope. The details of the instruments are described below. 

\subsubsection{Long-slit Intermediate Resolution Spectrograph for Astronomy (LISA)}
We obtained low-resolution spectroscopic observations on different dates using the LISA instrument. The major details are already mentioned in \citet{Ahuja_2025_MNRAS}. Both the comet and the spectrophotometric standards were observed with the same slit (1.76 arcsec width; 2 arcmin length). The orientation of the slit is in the North $-$ South direction. To reduce the datasets, we collected bias frames, flat frames to correct for pixel response, separate sky frames at 1$^{\circ}$ away from the comet photocenter in the direction of comet's motion for background subtraction from the comet, and ArNe (Argon-Neon) lamp frames to wavelength-calibrate the spectra. Multiple spectrophotometric standard stars from the ESO database\footnote{\url{https://www.eso.org/sci/observing/tools/standards/spectra/stanlis.html}} and solar stars, primarily HD19445 (G2V), were observed for flux calibration and continuum correction. The average seeing of during the observations from the LISA/PRL instrument was 2 arcsec. The spectroscopic observation details are mentioned in Table \ref{tab: observation_details}.

\subsubsection{Wide-Field Imager (WFI)}
For photometric observations, we used the WFI instrument mounted on the PRL 0.5$-$m and 1.2$-$m telescopes at different epochs. The main details of the instrument are covered in section 3.5 of \citet{Das_2026}; here, we briefly summarise them. The detector has a 11256 $\times$ 8348 pixel array with a pixel size of 3.76 microns. The comet was observed in the 4 $\times$ 4 binning mode, which yields an effective plate scale of 0.199 arcsec pixel $^ {-1}$ and a Field of View (hereafter FOV) of 9.3 arcmin $\times$ 7.6 arcmin when the instrument is mounted on the 1.2$-$m telescope. The FOV is 44 arcmin $\times$ 33 arcmin with a pixel scale of 0.91 arcsec pixel $^ {-1}$ on the 0.5$-$m telescope. The instrument is equipped with a set of Johnson filters. 

\subsubsection{Faint Object Camera (FOC)}
As the comet faded, we also used the FOC instrument mounted on the Cassegrain focus of the PRL 2.5$-$m telescope. The instrument consists of a deep-depletion back-illuminated ANDOR iKon-XL 231 CCD with a 4096 $\times$ 4108 pixel array of 15 micron pixel size.  The instrument is equipped with the Sloan filters. Images are taken in 2 $\times$ 2 binning mode, providing an effective instrument plate scale of 0.31 arcsec pixel$^ {-1}$ and a FOV of 10 arcmin $\times$ 10 arcmin.

\subsection{2-m Himalayan Chandra Telescope (HCT)}\label{HCT}
Photometric and spectroscopic observations were conducted using the Hanle Faint Object Spectrograph and Camera (HFOSC\footnote{\url{https://www.iiap.res.in/centers/iao/facilities/hct/hfosc/}}) instrument mounted at the Cassegrain focus of the 2$-$m, $f/9$, Ritchey-Chretien, HCT installed at the Indian Astronomical Observatory (IAO), Hanle, India (lat: 32.78$^\circ$ N; long 75.96$^\circ$ E; alt: 4500 m) \citep{HCT_paper}. Detailed instrumental and observational characteristics are described in section 2 of \citet{Ahuja_2025_MNRAS}.

For the photometric observations, the comet has been observed using the different Sloan filters installed in HFOSC. The FOV of the HFOSC is 10 arcmin $\times$ 10 arcmin, having a plate scale of 0.296 arcsec pixel$^{-1}$. The spectroscopic observations were carried out using Grism 7 (3800–6840\,\AA; R$\sim$1330) to cover the principal cometary emission bands. The comet was observed with the 167l slit (slit width: 1.92 arcsec, slit length: 11 arcmin), while spectrophotometric standards were obtained with the 1340l slit (slit width: 15.4 arcsec, slit length: 11 arcmin) to minimise slit losses. The slit orientation is East $-$ West. The slits used in LISA and HFOSC are oriented differently (see Fig. 1 of the \citet{Ahuja_2025_MNRAS}), sampling different directions in the cometary coma. Comet sky frames were acquired by offsetting the telescope by $1^\circ$ in declination in the direction of the comet's motion. The observed sky spectra were checked regularly to confirm that no contamination was coming from the comet's spectra.

Spectrophotometric standard stars from the ESO standards database were observed at airmasses comparable to the comet for flux calibration. The average seeing during the observations using the HFOSC/HCT instrument was 2.5 arcsec. The solar analogue HD19445 \citep{HD_1918} was observed to remove the solar continuum. Bias, flat-field, and FeAr lamp frames were obtained following the standard HFOSC calibration procedure \citep{Ahuja_2025_MNRAS}.

\subsection{0.132-m Cepheid Observatory (CepO)}
CepO hosts a small-aperture telescope with a 0.132$-$m diameter, $f/7$, installed at (Lat: 24.55 deg N; Long: 75.34 deg E; Alt: 415 m). The native focal length of the telescope, 910 mm, was reduced to 665 mm by using a 0.7x reducer lens placed between the telescope objective lens and the CCD sensor. This increases the FOV and includes more photometric standard stars within the same air mass, resulting in a $f/5.03$ beam. The photometric observations were carried out using an Atik383L+ CCD with a 3354 $\times$ 2529 pixel chip, each pixel measuring 5.4 microns. To enhance sensitivity, the images are binned into 2$\times$2. The effective plate scale is 3.34 arcsec pixel$^{-1}$. The observations were carried out from 23 to 29 December 2025 using an $R$-band filter and one epoch in $B$ and $V$ filters on 31 December 2025. The observations were taken at geocentric distances ranging from 1.80 au to 1.87 au.

\section{Data Reduction Methods and Analysis}\label{sec: data reduction}
\subsection{Spectroscopic data reduction}
Spectroscopic data from different instruments, i.e.,  HFOSC/HCT and LISA/PRL, were reduced using the procedures described in \citet{Ahuja_2025_MNRAS}. Briefly, the data reduction includes bias subtraction, flat-fielding, cosmic-ray removal using the LA Cosmic package \citep{LACosmic} and one-dimensional spectral extraction using self-developed \textsc{Python} codes. The wavelength calibration is performed using an arc lamp spectrum, and the flux calibration is performed using spectrophotometric standards with standard \textsc{iraf} packages. Solar continuum removal from the comet spectrum is done by observing the solar analogue (G2V) stars, mainly HD19445, following procedures described in \citet{continuum_removal}. We use the continuum windows mentioned in \citet{continuum_removal} and fit third or fifth-order polynomials to the flux-calibrated comet and solar spectra. The solar spectrum was normalised by the polynomial-fitted solar spectrum. The resulting normalised spectrum was multiplied by the polynomial fit of the continuum obtained from the comet spectrum.  This represents the synthetic dust continuum in the comet and is subtracted from the comet spectrum to get continuum-corrected spectra for the comet.

Since the slit used in the LISA/PRL spectrograph has a width of 1.76 arcsec, which is narrower than the typical seeing during most nights. This would result in the slit loss of standard star flux, affecting the flux calibration of the comet. Hence, to correct for it, we used the method given by \citet{Lee_Pak}, in which the slit correction factor is calculated as a function of wavelength using properties of the error function. This correction factor was multiplied by the comet's flux-calibrated spectrum. Also, the comet spectrum is further corrected for slit loss due to atmospheric dispersion of the standard star counts, which varies with wavelength \citep{Filippenko_1982,Stone_1996}. The continuum-corrected spectra for comet 3I, taken using LISA and HFOSC instruments, are shown in Fig. \ref{spec_3I}.

\subsection{Photometric Data Reduction}
For the calibration of the photometric datasets obtained from the 0.132$-$m CepO, bias and flat frames were applied to correct the datasets, and the absolute photometry was performed with the Tycho Tracker Software \citep{Parrott_2020,2025epsc.conf.1222P}. The software uses the Gaia DR3 for astrometry \citep{Gaiadr3_1} and the ATLAS catalogue for photometric reference \citep{ATLAS_1}. The Tycho software also uses the transformation equations given by \citet{Panstarrs_1} to derive the Johnson-Cousins $BVRI$ magnitudes. Aperture photometry with a 3-pixel radius was applied in the software, and the magnitude of the comet in the $R$-band filter was automatically calibrated by selecting the catalog of standard stars in the observed field.

For datasets obtained with other telescopes, we used custom \textsc{Python} routines to reduce the comet and field star frames. The photometric calibration was performed using the \textsc{Photutils} tool from the \textsc{Astropy} package (\citet{Astropy_2013}). Apertures of 2-3 $\times$ Full Width Half Maximum (FWHM) were used to measure the magnitudes of the field stars and determine the photometric zero point, which was then applied to compute the comet's magnitude for an aperture size of 10000 km. The reference magnitudes for field stars were obtained from the Pan-STARRS \citep{Chambers2016} catalog for data acquired with the 2.5$-$m telescope, whereas for the 0.5 $-$ and 1.2$-$m class telescopes, the magnitudes were retrieved from the SIMBAD database \citep{Wenger2000} and vizieR catalogues \citep{vizier}.

\subsection{\texorpdfstring{Gas production rates and dust proxy parameter Af$\rho$}{Gas production rates and dust proxy parameter Afrho}}\label{Gas_dust}

The molecular gas production rates of CN (0$-$0), C$_2$ ($\Delta=$0), and C$_3(0-0)$ are derived by calculating the column density profiles over different radial apertures. They are then compared with the theoretical column density profiles from the Haser model \citep{Haser_1957} \citep[see sections 4.1 and 4.2 of][]{Ahuja_2025_MNRAS}. The parent-daughter scale lengths for the respective molecules are adopted from \citet{AHEARN_1995}. The molecular bandpasses are taken from Table 5 of \citet{Langland-Shula2011}, and the fluorescence efficiency ($g$) factors for C$_2$ and for C$_3$ molecules are adopted from Table 2 of \citet{AHEARN_1995}. For the CN molecule, whose  $g$ factor is affected by the Swings effect \citep{Swings_1941}, the method given by \citet{Schleicher_2010} are used to calculate the $g$-factor using the double interpolation of the heliocentric distance and velocity. These details are compiled in the Table \ref{tab: haser_parameters}.

\begin{table}
\centering
\normalsize
\caption{Haser-model parent and daughter scale lengths at 1 au adopted from \citet{AHEARN_1995}. Fluorescence efficiencies are given at 1 au.}
\label{tab: haser_parameters}

\begin{tabular}{lccc}
\hline
& \multicolumn{2}{c}{Scale Length ($10^4$ km)} & \\
\cline{2-3}
Species & Parent & Daughter & $g(1\,{\rm au})$ \\
\hline
CN      & 1.3 & 21.0 & varies$^{a}$ \\
C$_2$   & 2.2 & 6.6  & $4.5\times10^{-13}$ \\
C$_3$   & 0.28 & 2.7  & $1.0\times10^{-12}$ \\
\hline
\end{tabular}

\vspace{0.15cm}

\footnotesize
$^{a}$ The CN fluorescence efficiency depends on heliocentric distance and radial velocity due to the Swings effect \citep{Swings_1941}. Hence, it is calculated following the method by \citet{Schleicher_2010}.

\end{table}

The scale lengths vary directly with the heliocentric distance ($L \propto r_h^{2}$), whereas the fluorescence efficiency factor varies in an inverse relationship with the heliocentric distance ($g \propto r_h^{-2}$). The gas outflow velocity is assumed to be 1 km s$^{-1}$. 

For faint spectra, removing the continuum is very difficult. Therefore, the single-aperture of particular arcsec is extracted and multiplied by the Haser correction \citep[See][and references therein]{Ahuja_2025_MNRAS}. The Haser fraction (inverse of the Haser correction) can be calculated for the respective heliocentric distance, geocentric distance and aperture radius\footnote{\url{https://asteroid.lowell.edu/comet/haser}}.

\begin{figure*}
    \centering
    \includegraphics[width=\linewidth]{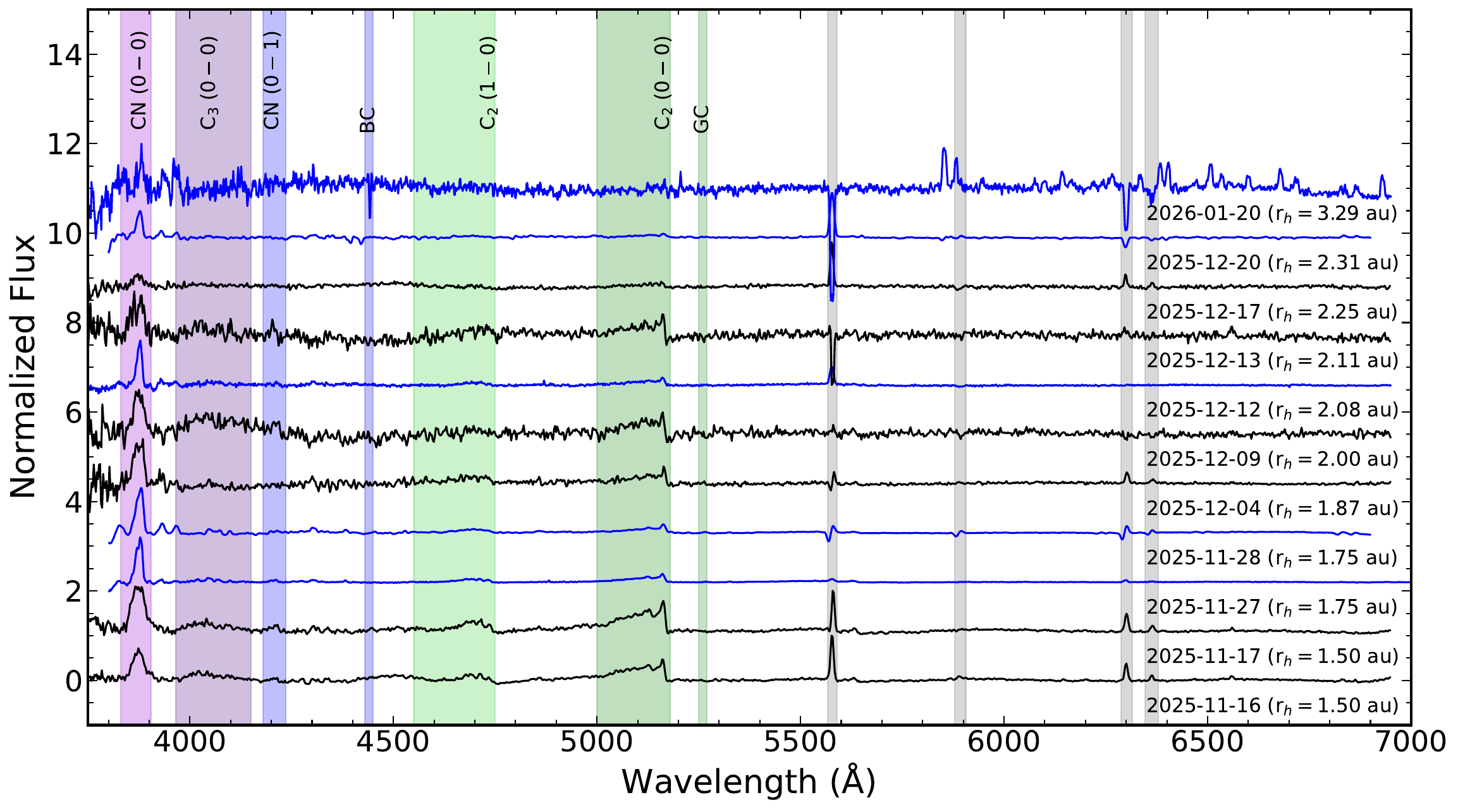}
    \caption{Solar continuum corrected spectra of comet 3I observed from LISA/PRL (in black) and HFOSC/HCT (in blue). The spectra are normalised to the CN $(0$-$0)$ emission band and spaced by 1 relative flux unit for clarity. Prominent molecular emission bands are marked. The BC and GC continuum windows used for the Af$\rho$ measurements are also indicated. Regions affected by strong sky emissions are shown with gray shaded vertical bands.}
    \label{spec_3I}
\end{figure*}

The cometary dust activity is measured using the dust proxy parameter, Af$\rho$ \citep{Ahearn_1984}  in the Blue-Continuum (hereafter BC) and Green-Continuum (hereafter GC) bands. Both regions are illustrated in the Fig. \ref{spec_3I}. The central wavelengths for BC and GC are 4453 \AA\ and 5259 \AA, respectively. These continuum bandpasses are taken from \citet{Farnham_1998}. The details for computing the parameter from long-slit spectroscopy are given in section 4.3 of \citet{Ahuja_2025_MNRAS}. 

\section{Results and Discussion}\label{sec: Results}
In this section, we will provide the results obtained from the spectroscopic and photometric observations of comet 3I.

\subsection{Molecular Production Rates, Power Law Fit, and Abundance Ratios}\label{sec: Results_1}
\begin{table*}
\caption{Spectroscopic gas production rates and Af$\rho$ values derived from Blue-Continuum (BC) and Green-Continuum (GC) passbands using observations from multiple telescopes.}
\setlength{\tabcolsep}{3.0pt}
        \resizebox{\textwidth}{!}{%
\begin{tabular}{lcccccccccccc}
\hline
\multicolumn{1}{|c|}{{Date}} &
\multicolumn{1}{c|}{{Instrument/}}&
  \multicolumn{1}{c|}{{TFP}} &
  \multicolumn{1}{c|}{{r$_h$}} &
  \multicolumn{1}{c|}{{$\Delta$}} &
  \multicolumn{3}{c|}{\hspace{0.5cm}{Production Rates ($\times$ 10$^{24}$ molecules s$^{-1}$)}} &
  \multicolumn{2}{c|}{{A($\theta=0^{\circ}$)f$\rho$ (cm)}}\\
\multicolumn{1}{|c|}{} &
\multicolumn{1}{c|}{{Telescope}}&
  \multicolumn{1}{c|}{{(Days)}} &
  \multicolumn{1}{c|}{{(au)}} &
  \multicolumn{1}{c|}{{(au)}} &
  \multicolumn{1}{c|}{{Q(CN)}} &
  \multicolumn{1}{c|}{{Q(C\textsubscript{2})}} &
  \multicolumn{1}{c|}{{Q(C\textsubscript{3})}} &
  \multicolumn{1}{c|}{{BC}} & 
  \multicolumn{1}{c|}{{GC}}\\
  \hline
2025 Nov 16 & LISA / 1.2$-$m PRL & 17.5 & 1.50 & 2.09 & 80.72 $\pm$ 3.99 & 69.92 $\pm$ 2.35 & 10.68 $\pm$ 1.96 & 2036.45 $\pm$ 353.87 & 3133.67 $\pm$ 201.66\\
2025 Nov 17 &  LISA / 1.2$-$m PRL & 18.5 & 1.50 & 2.09 & 89.87 $\pm$ 3.70 & 63.36 $\pm$ 2.88 & 10.76 $\pm$ 2.29 & -- & --\\
2025 Nov 27 &  HFOSC / 2$-$m HCT & 28.5 & 1.75 & 1.95 & 55.40 $\pm$ 0.51 & 56.20 $\pm$ 3.29 & 6.79 $\pm$ 0.77 & 2552.43 $\pm$ 230.67 & 3298.70 $\pm$ 158.27\\
2025 Nov 28 &  HFOSC / 2$-$m HCT & 29.5 & 1.75 & 1.95 & 58.80 $\pm$ 0.95 & 55.40 $\pm$ 5.15 & 6.21 $\pm$ 1.22 & 2739.77 $\pm$ 273.14 & 3352.40 $\pm$ 206.31\\
2025 Dec 04 &  LISA / 1.2$-$m PRL & 35.5 & 1.87 & 1.88 & 41.13 $\pm$ 2.01 & 25.43 $\pm$ 5.32 & 3.25 $\pm$ 1.42 & 2719.96 $\pm$ 517.11 & 3335.17 $\pm$ 368.94\\
2025 Dec 09 &  LISA / 1.2$-$m PRL & 40.5 & 2.00 & 1.84 & 37.60 $\pm$ 9.16 & 22.88 $\pm$ 4.53 & -- & 2985.86 $\pm$ 567.66 & 3661.21 $\pm$ 405.01\\
2025 Dec 12 &  HFOSC / 2$-$m HCT & 43.5 & 2.08 & 1.82 & 22.26 $\pm$ 1.39 & 20.54 $\pm$ 2.92 & 4.35 $\pm$ 2.40 & 908.34 $\pm$ 297.14 & 1124.81 $\pm$ 194.34\\
2025 Dec 13 &  LISA / 1.2$-$m PRL & 44.5 & 2.11 & 1.81 & 21.60 $\pm$ 2.76 & 21.67 $\pm$ 4.76 & 6.04 $\pm$ 2.65 & 1957.36 $\pm$ 568.18 & 2317.68 $\pm$ 285.47\\
2025 Dec 17 &  LISA / 1.2$-$m PRL & 48.5 & 2.25 & 1.80 & 17.06 $\pm$ 3.11 & 16.93 $\pm$ 9.69 & 5.01 $\pm$ 3.06 & 2010.19 $\pm$ 568.86 & 2637.73 $\pm$ 453.05\\
2025 Dec 20 &  HFOSC / 2$-$m HCT & 43.5 & 2.31 & 1.80 & 18.55 $\pm$ 2.32 & 10.63 $\pm$ 3.11 & 3.03 $\pm$ 2.73 & 595.54 $\pm$ 166.89 & 672.46 $\pm$ 120.23\\
2026 Jan 20 &  HFOSC / 2$-$m HCT & 83.5 & 3.29 & 2.31 & 13.50 $\pm$ 5.89 & 10.10 $\pm$ 9.14 & -- & 741.06 $\pm$ 332.16 & 1013.85 $\pm$ 195.94\\
\hline
\end{tabular}
}
\label{tab: Production rate- I}
\end{table*}
After obtaining the continuum-corrected spectra following the reduction steps described in section \ref{sec: data reduction}, we derived the production rates of various molecules, CN $(0$-$0)$, C$_2 (\Delta$=$0)$, and C$_3(0$-$0)$ using the methodology described in section \ref{Gas_dust}. The derived production rates are listed in Table \ref{tab: Production rate- I}. For comparison, we also considered other available post-perihelion production rates for comet 3I reported by \citet{Jehin_17515_2025,Jehin_17538_2025}, \citet{Hoogendam_2026}, and \citet{Zhao_2026}.

\subsubsection{Molecular Production Rates}
The evolution of different molecular production rates as a function of heliocentric distance and the number of days from perihelion is presented in Appendix \ref{Qs}. As shown in Fig. \ref{QCN_fit}, \ref{QC2_fit}, and \ref{QC3_fit}, we compare our measured production rates with those reported by \citet{Jehin_17515_2025,Jehin_17538_2025}, \citet{Hoogendam_2026} and \citet{Zhao_2026}. For consistency, all production rates are converted to an outflow velocity of 1 km s$^{-1}$ before comparison. The values provided by \citet{Ganesh_17502_ATel_2025} have been further corrected to account for the effect of high airmass on the sensitivity function used in the production rates calculation. Since \citet{Zhao_2026} adopted different Haser parameters by \citet{Cochran_2012}, their production rates are approximately re-scaled based on the differences in the adopted fluorescence efficiencies, gas outflow velocities, and parent–daughter scale lengths given by A'Hearn \citep{AHEARN_1995}. However, for the abundance-ratio analysis, we have used the published values reported by \citet{Zhao_2026}.
As seen, the production rates of CN (Fig. \ref{QCN_fit}) and C$_2$ (Fig. \ref{QC2_fit}) are in close agreement with all included datasets. However, our C$_3$ production rate (Fig. \ref{QC3_fit}) is in the range measured by \citet{Jehin_17538_2025} and even within the errorbar for the values observed by \citet{Zhao_2026}. However, it does not match the values observed by \citet{Hoogendam_2026}.

\subsubsection{Power Law Fit}
To quantify the variation of CN, C$_2$ and C$_3$ with heliocentric distance, weighted power-law fits using the \texttt{curve\_fit} package from \textsc{Python} were applied to the different molecular species by considering only our datasets and by including all the available datasets, as shown in Figs. \ref{QCN_fit}, \ref{QC2_fit}, and \ref{QC3_fit}. The fitted parameters are given in Table \ref{tab: Power Law}. The derived indices for CN and C$_2$ are consistent with those reported by \citet{Hoogendam_2026} and \citet{Zhao_2026} when only our datasets are considered. However, the number changes significantly for CN when other datasets are included in the fit. For the C$_3$ molecule, the fit differs, possibly due to the large variation in measured production rates. We also would like to mention that both CN and C$_3$ molecular bands are contaminated by Fe\, I lines (See Fig. 1 and 2 of \citealp{Hoogendam_2026}). This would affect the derived production rates. High-resolution spectroscopy could help eliminate the effects of metallic lines and provide precise production rates. 

\subsubsection{Production Rate Ratios}
\begin{figure}
    \centering
    \includegraphics[width=1.0\linewidth]{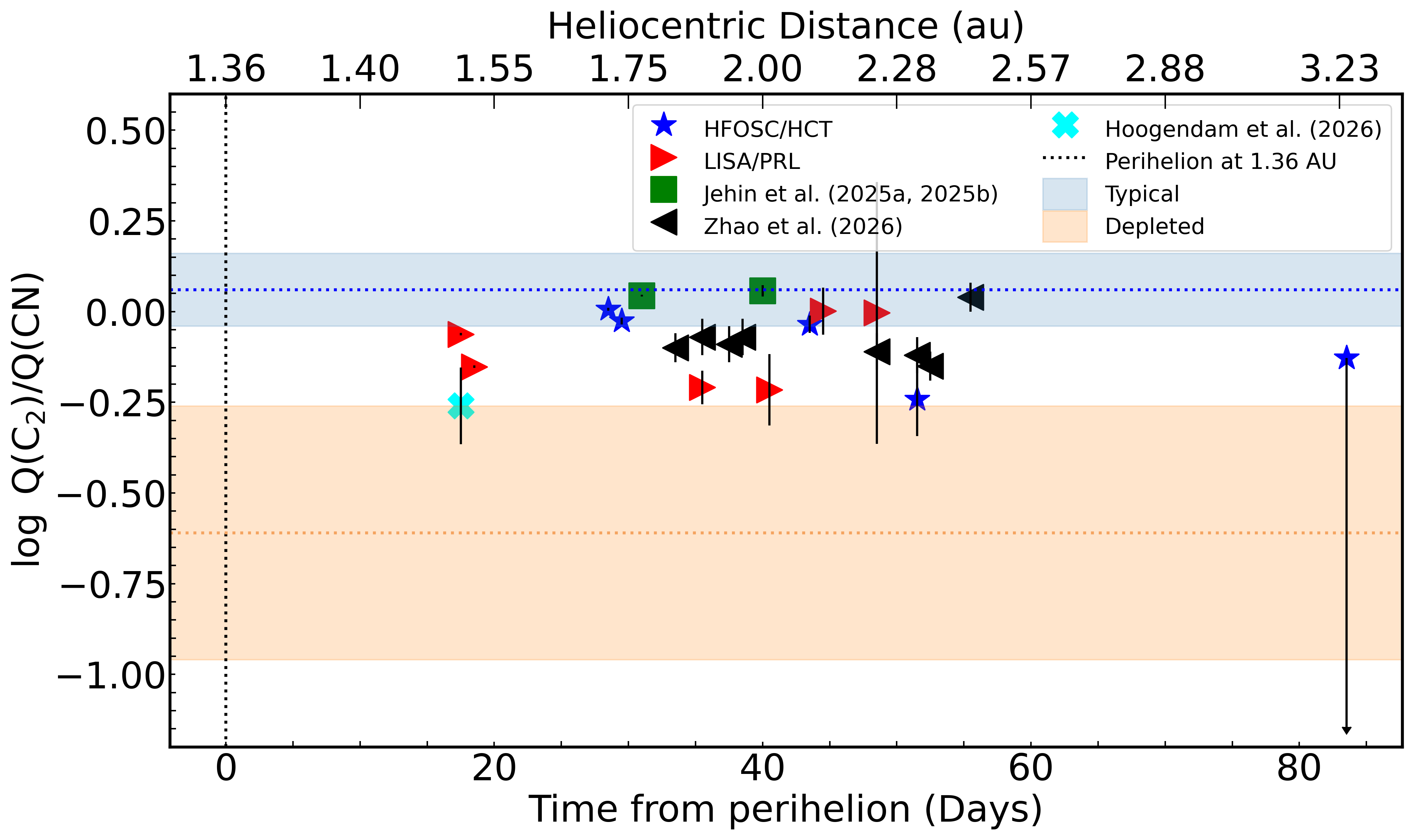}
    \caption{Variation of C$_2$/CN with the heliocentric distance. We have marked the 1-$\sigma$ range of the typical region with light blue and the 1-$\sigma$ range of the depleted region with orange. The observation from LISA/PRL is marked as red right triangle, from HFOSC/HCT is marked as blue asterisk, from \citet{Jehin_17515_2025,Jehin_17538_2025} marked as green square, from \citet{Hoogendam_2026} marked as cyan cross, and the observation from \citet{Zhao_2026} marked as black left triangle. The vertical dotted line, marked in this figure (and all others), represents the perihelion distance, which is 1.36 au.}
    \label{C2_CN}
\end{figure}

We then calculated the production rate ratios, namely $\log~[Q(\mathrm{C}_2)/Q(\mathrm{CN})]$ and $\log~[Q(\mathrm{C}_3)/Q(\mathrm{CN})]$ as well as dust-to-gas ratio, $\log~[Af\rho/Q(\mathrm{CN})]$. Production rate ratios are commonly used to classify cometary compositions as carbon-typical or carbon-depleted, following the criteria defined by \citet{AHEARN_1995}. As shown in Fig. \ref{C2_CN}, the post-perihelion $\log~[Q(\mathrm{C}_2)/Q(\mathrm{CN})]$ ratios show a significant increase compared to the pre-perihelion values, with several measurements now lying within the typical region (\textit{light blue} region). The derived values are in good agreement with the numbers reported by \citet{Jehin_17515_2025,Jehin_17538_2025}, \citet{Hoogendam_2026} and \citet{Zhao_2026}. The mean value obtained from our datasets is $-0.06 \pm 0.09$, which has substantially changed relative to the pre-perihelion values of approximately $-0.8$ reported at a heliocentric distance of around $\sim$2.8 au by \citet{Salazar_Manzano_2025} and \citet{Lazzarin_2026}.

\begin{figure}
    \centering
    \includegraphics[width=1.0\linewidth]{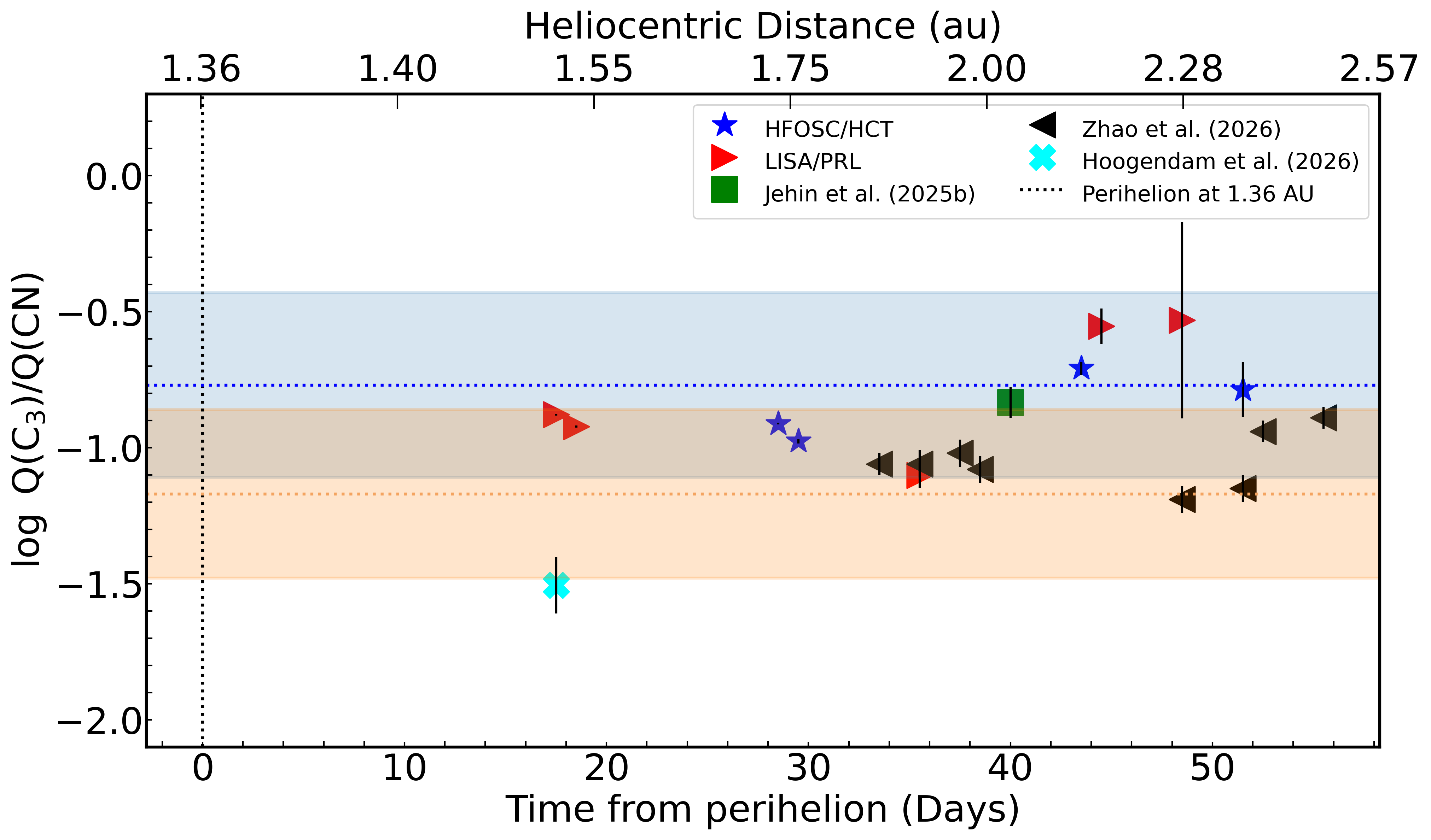}
    \caption{Variation of C$_3$/CN with the heliocentric distance. We have marked the 1-$\sigma$ range of the typical region with light blue and the 1-$\sigma$ range of the depleted region with orange. The observation from LISA/PRL is marked as a red right triangle, from HFOSC/HCT is marked as a blue asterisk, from \citet{Jehin_17538_2025} marked as a green square, from \citet{Hoogendam_2026} marked as a cyan cross, and the observation from \citet{Zhao_2026} marked as a black left triangle.}
    \label{C3_CN}
\end{figure}

Similarly, the $\log~[Q(\mathrm{C}_3)/Q(\mathrm{CN})]$ ratio shown in Fig. \ref{C3_CN} exhibit a similar trend, indicating a significant increase relative to the pre-perihelion measurements. The observed values are similar to those reported by \citet{Jehin_17538_2025} and \citet{Zhao_2026}. whereas \citet{Hoogendam_2026} still reported values within the depleted region. The mean value of $\log~[Q(\mathrm{C}_3)/Q(\mathrm{CN})]$ derived from our observed values is $-0.88 \pm 0.18$, which has again increased from $-1.47$ reported by \citet{Lazzarin_2026}. Hence, these results confirm the transition of the composition of comet 3I from carbon-depleted to carbon-typical. 

This change in the chemical composition of comet 3I may be attributed to Galactic Cosmic Ray (GCR) processing of the surface layers of the comet \citep{Maggiolo_2026}, which could have affected the volatile composition observed during the pre-perihelion phase. Following perihelion passage, the exposure of the less-processed inner layers is likely to explain the observed shift in composition, which is consistent with the findings of \citet{Lisse_2026} and \citet{Belyakov_2026}. A similar behaviour was observed in the second interstellar comet 2I/Borisov, where the post-perihelion $\log~[Q(\mathrm{C}_2)/Q(\mathrm{CN})]$ ratio increased relative to its pre-perihelion values, although the comet 2I/Borisov still remained within the carbon-depleted region (see Fig. 4 of \citet{Aravind_2I_2021}). This may be due to its relatively distant perihelion of $\sim$2.01 au, which may not have been sufficient to completely remove the outer processed layers.

\subsection{\texorpdfstring{Dust proxy parameters Af$\rho$ and Dust-to-Gas Ratio}{Dust proxy parameters Afrho and Dust-to-Gas Ratio}}\label{sec: Results_2}
We also calculated the dust proxy parameter in the Blue Continuum (BC) region (Af$\rho_{\rm BC}$) and the Green Continuum (GC) region (Af$\rho_{\rm GC}$), and the corresponding values are listed in Table \ref{tab: Production rate- I}. As can be seen in Fig. \ref{A_BC} and \ref{A_GC}, most of the estimated Af$\rho$ values are consistent with the values reported by \citet{Jehin_17515_2025,Jehin_17538_2025}, though a few points exhibit noticeable dispersion in the estimated values. The large range in the dispersion could be due to the orientation of the slit sampling regions with strong non-uniformity. To investigate this dispersion, we analysed the Sloan $r$-band images obtained on 2025 Dec 20 using the modified Larson–Sekanina (LS) processing (see section 6.4.2 of \citet{Ahuja_2025_MNRAS} and the references therein). The processed images revealed two jet-like features emanating from the coma (see Fig. \ref{fig: LS_afp}). We found that the jet structures were misaligned with the orientation of the HFOSC/HCT slit, which likely affected the Af$\rho$ measurements in the BC and GC regions. This suggests that the Af$\rho$ should be estimated from 2-D data rather than a one-dimensional slit in the case of an asymmetric coma.

\begin{figure}
    \centering
    \includegraphics[width=1.0\linewidth]{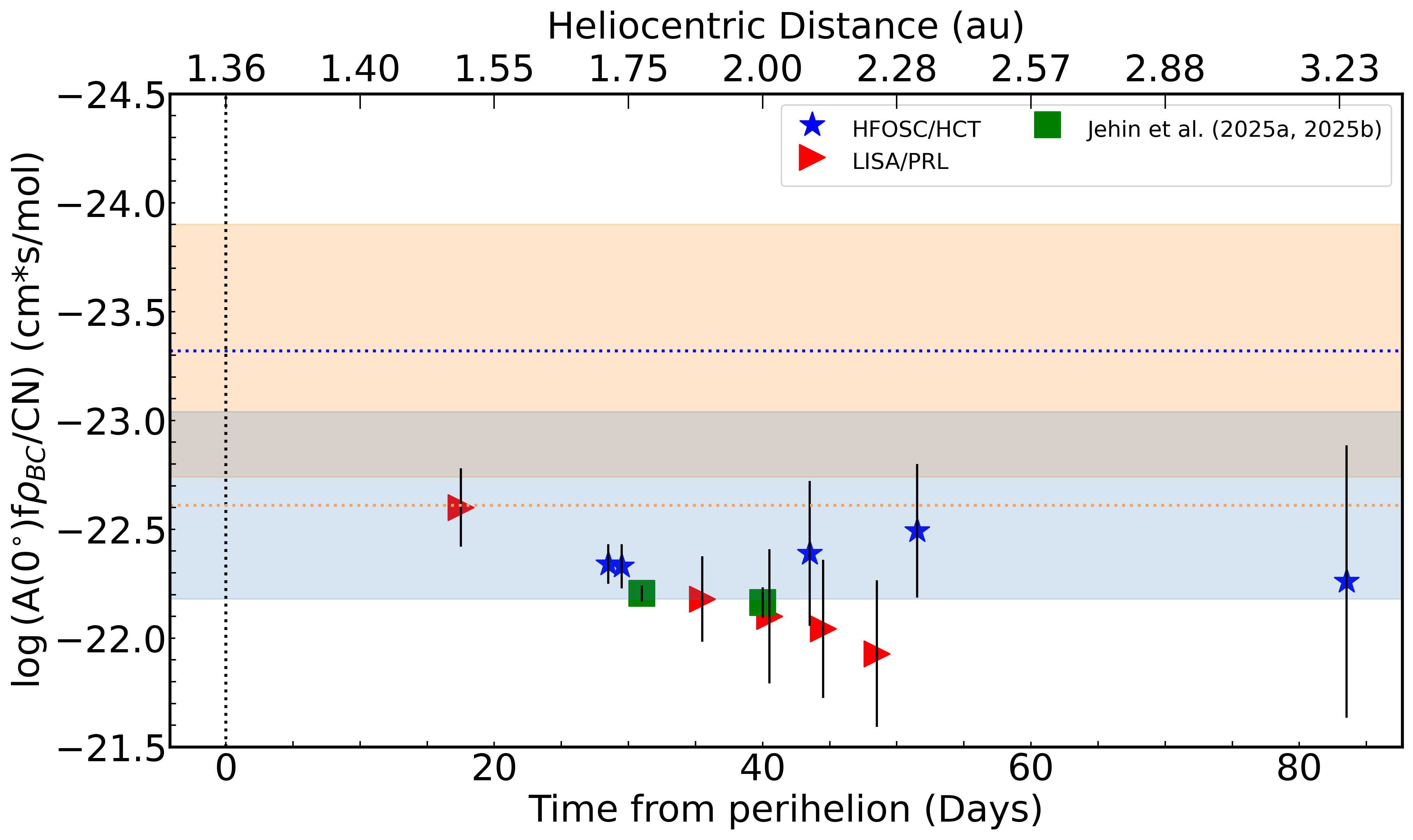}
    \caption{Variation of Af$\rho_{BC}$/CN with the heliocentric distance. We have marked the 1-$\sigma$ range of the dust-rich region with light blue and the 1-$\sigma$ range of the dust-poor region with orange. The observation from \citet{Jehin_17515_2025,Jehin_17538_2025} is marked as a green square, the observation from HFOSC/HCT is marked as a blue asterisk, and the observation from LISA/PRL is marked as a red right triangle.}
    \label{Afp_CN}
\end{figure}

We then examined the dust-to-gas ratio, Af$\rho_{\rm BC}$/CN, a key parameter to distinguish whether the comet is dust-rich or dust-poor \citep{AHEARN_1995}. As shown in Fig. \ref{Afp_CN}, all the measurements are in the dust-rich region, with a mean value of Af$\rho_{BC}$/CN of -22.30 $\pm$ 0.20. Hence, the interstellar comet 3I is confirmed to be a dust-rich comet. 

\subsection{Broadband Colours}\label{sec: Results_3}
\begin{figure}
    \centering
    \includegraphics[width= \linewidth]{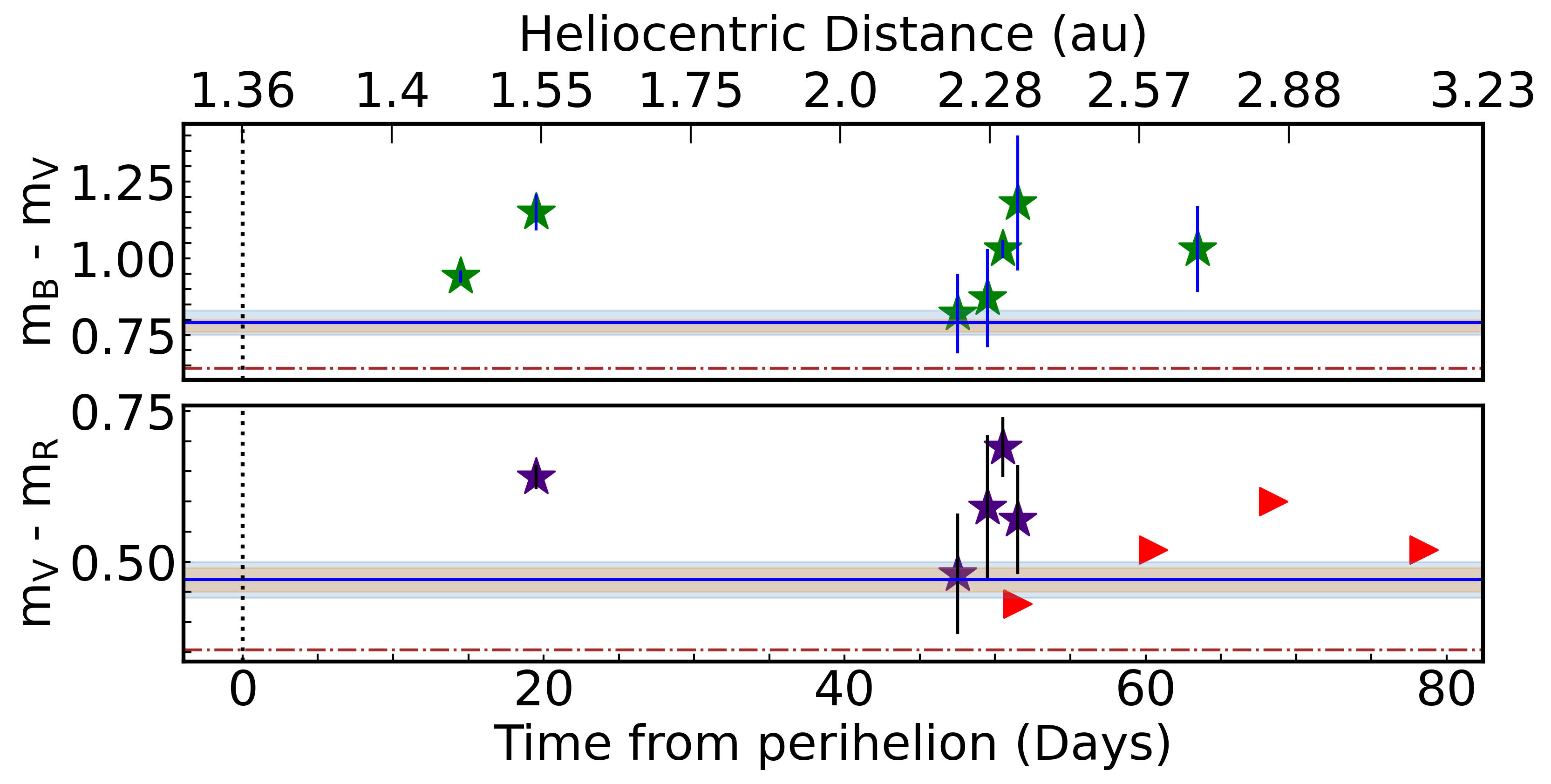}
        \caption{Variation of the colour indices $B-V$ (above, green asterisk), $V-R$ (below, indigo asterisk) and $g-r$ (below, red right triangle) with days to perihelion and heliocentric distance (in au). The 1-$\sigma$ range for DNCs \citep{kulyk_2018} is marked as blue and for LPCs \citep{Jewitt_2015} as orange. The solar colours are marked as a dash-dot line in dark-red. The blue line is the median value of the DNCs \citep{kulyk_2018}.}
    \label{colour1}
\end{figure}

\begin{figure}
    \centering
    \includegraphics[width= \linewidth]{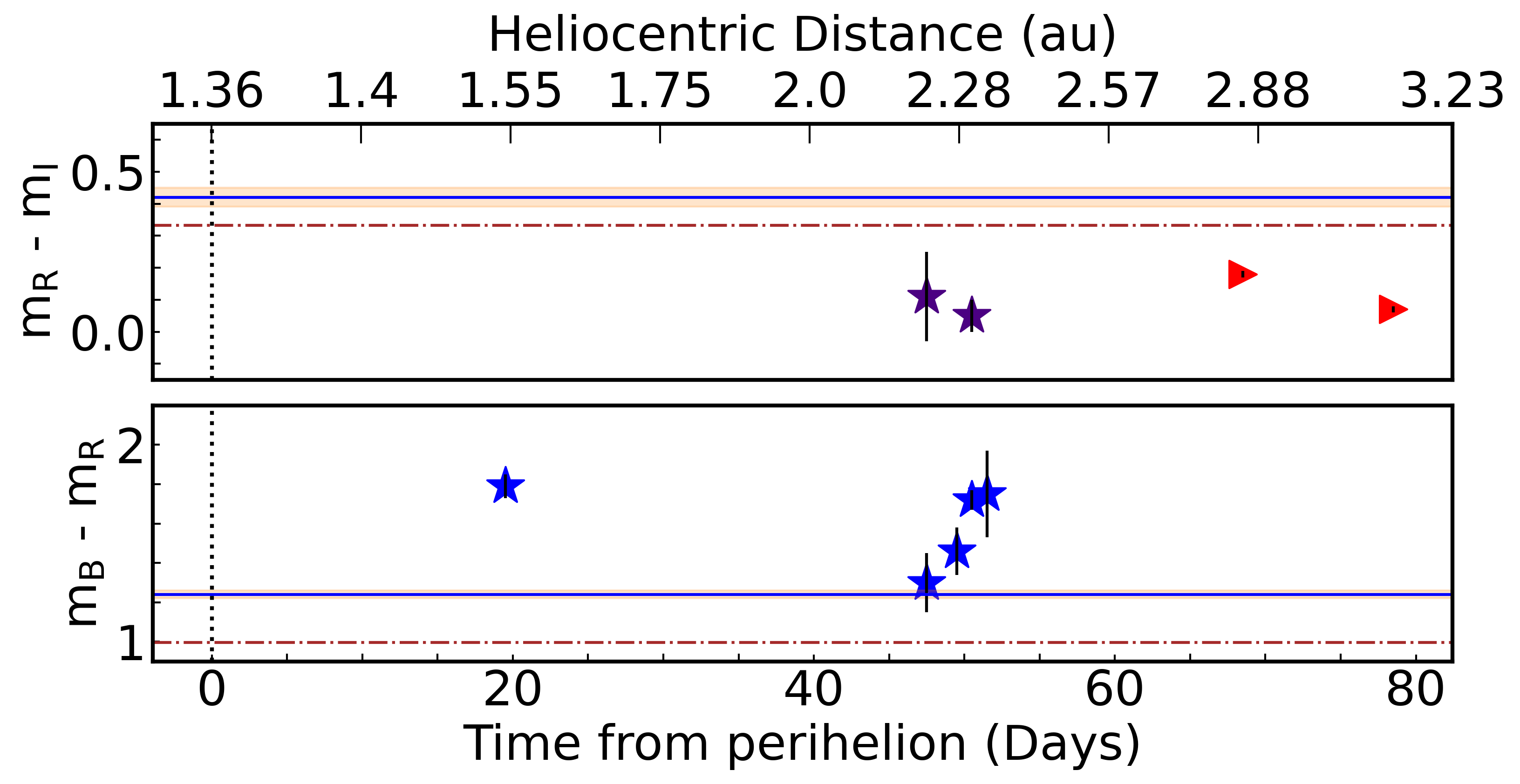}
    \caption{Variation of the colour indices $R-I$ (above, indigo star), $r-i$ (above, red right triangle) and $B-R$ (below, blue star) with days to perihelion and heliocentric distance (in au). The 1-$\sigma$ range for LPCs \citep{Jewitt_2015} is marked as orange. The solar colours are marked as a dash-dot line in dark-red. The blue line is the median value of the LPCs \citep{Jewitt_2015}.}
    \label{colour2}
\end{figure}

Using the results from the photometric observations shown in Tables \ref{tab: Photometric PRL-I} and \ref{tab: Photometric PRL-II}, the median post-perihelion comet colours are $B-V =$ 1.03 $\pm$ 0.14 mag, $V-R =$ 0.59 $\pm$ 0.08 mag,  $R-I =$ 0.08 $\pm$ 0.04 mag, $g-r =$ 0.52 $\pm$ 0.07 mag, $r-i =$ 0.12 $\pm$ 0.08 mag. The variation in the different photometric colours is shown in Fig. \ref{colour1} and \ref{colour2}. The comet 3I broadband colours are redder than solar colours and the Dynamically New Comets (DNCs) and Long-Period Comets (LPCs) in $B-V$ and $V-R$, while it is relatively bluer in $R-I$. The post-perihelion $B-V$ and $V-R$ colours are found to be similar to the pre-perihelion colours observed by \citet{Opitom_3I_disc} and \citet{3I_disc_1}. The post-perihelion $R-I$ colour is matching with the pre-perihelion $R-I$ colour observed by \citet{3I_disc_1}. Also, as the comet 3I was observed in Sloan colours, the values $g-r$ and $r-i$ are within 1-$\sigma$ of the observed colours by \citet{Beniyama_2025} and slightly bluer than the pre-perihelion colours observed by \citet{Hoogendam_2025d}.

In order to see whether the presence of molecular gas emissions affects the broadband colours, we have calculated the magnitude of the flux-calibrated spectrum of the comet and its continuum spectrum separately using the \textsc{Pyphot} package (\citet{filter_2020,zenodopyphot}). We found that the magnitudes of both the flux-calibrated spectra and their continuum are similar. Hence, it confirms that there is no effect of molecular emissions on the cometary colours.

\begin{figure}
    \centering
    \includegraphics[width=\linewidth]{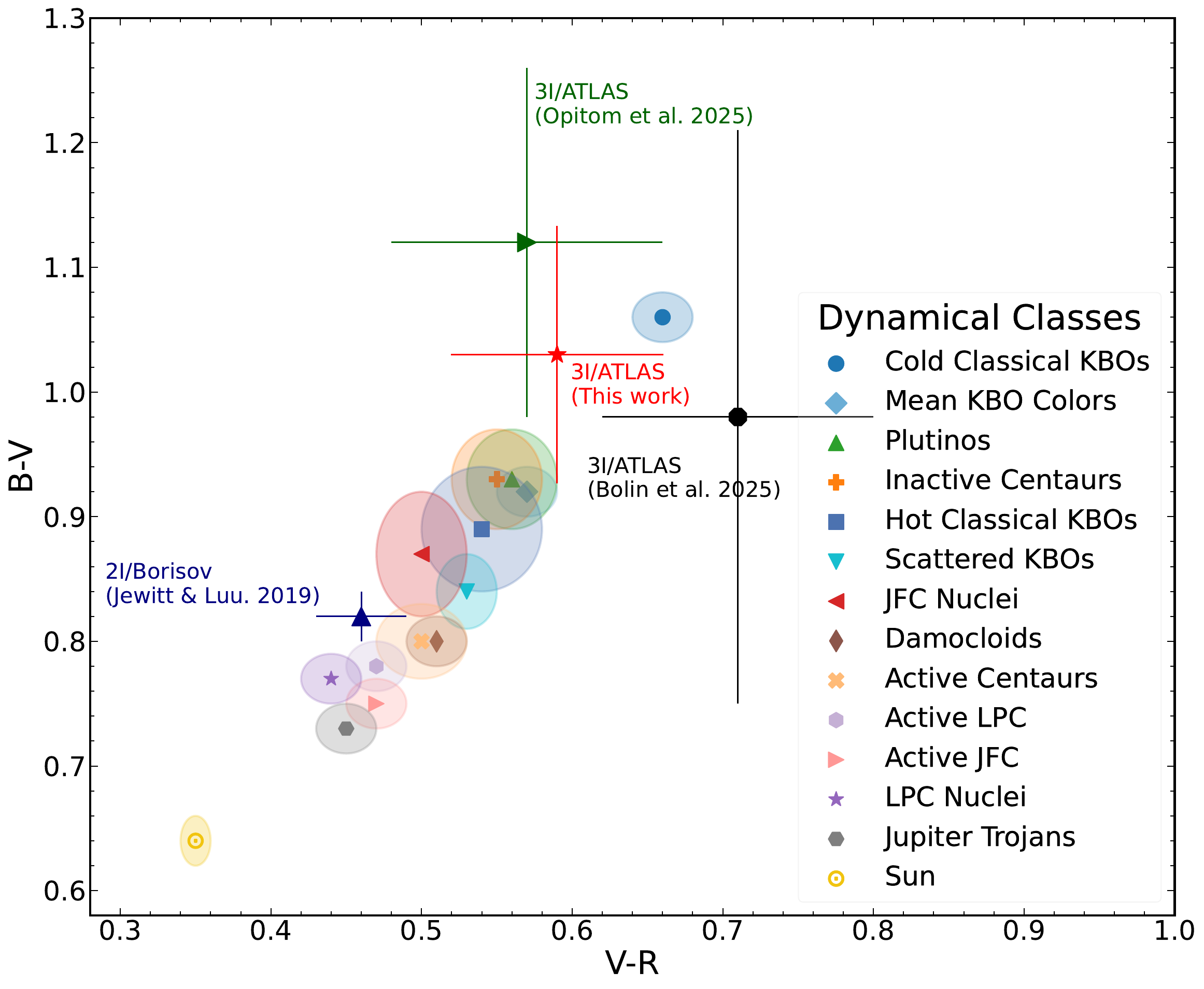}
    \caption{Colour – colour plot of $B-V$ versus $V-R$ for different dynamical classes of objects, with data compiled from \citet{Jewitt_2015}. The symbols represent the median colours, while the shaded ellipses (or circles) denote the $\pm1\sigma$ standard errors of the mean colours for each dynamical class. The pre-perihelion colour of comet 3I from \citet{3I_disc_1} is shown as a black circle, while the measurement from \citet{Opitom_3I_disc} is represented by a green right-pointing triangle. The colour of 2I/Borisov from \citet{Jewitt_2019} is also shown as a navy upward triangle. The solar colour is indicated by a yellow symbol, while the median post-perihelion colour of comet 3I is represented by a red star. The error bars correspond to the 1$\sigma$ standard deviation of the measured post-perihelion colours.}
    \label{fig: colour3I}
\end{figure}

Fig. \ref{fig: colour3I} shows the colour-colour diagram of various classes of minor bodies in the solar system \citep{Jewitt_2015}.  We have added our post-perihelion median colours of 3I and compared with the pre-perihelion median values from \citep{3I_disc_1,Opitom_3I_disc}. Also shown is the pre-perihelion colour of 2I/Borisov \citep{Jewitt_2019}.  It is found that the comet colours are close to the Cold classical KBOs.

\begin{table*}
\caption{Broad-band Bessel photometric observations obtained with WFI (on 0.5$-$m and 1.2$-$m). The reported magnitudes are median values of each dataset.}
\begin{threeparttable}
\setlength{\tabcolsep}{5.5pt}
        \resizebox{\textwidth}{!}{%
\begin{tabular}{ccccccccccc}
\hline
\multicolumn{1}{|c|}{{Date}} &
  \multicolumn{1}{c}{{Instrument/}} &
  \multicolumn{1}{c|}{{TFP}} &
  \multicolumn{1}{c|}{{r$_h$}} &
  \multicolumn{1}{c|}{{$\Delta$}} &
  \multicolumn{1}{c|}{{Airmass}} &
  \multicolumn{1}{c|}{{Phase}} &
  \multicolumn{4}{c|}{\hspace{0.5cm}{Median Magnitudes (mag)}}\\
\multicolumn{1}{c|}{} &
\multicolumn{1}{c|}{Telescope} &
  \multicolumn{1}{c|}{{(Days)}} &
  \multicolumn{1}{c|}{{(au)}} &
  \multicolumn{1}{c|}{{(au)}} &
  \multicolumn{1}{c|}{{}} &
  \multicolumn{1}{c|}{{Angle ($^{\circ}$)}} &
  \multicolumn{1}{c|}{{\textit{B}}} &
  \multicolumn{1}{c|}{{\textit{V}}} &
  \multicolumn{1}{c|}{{\textit{R}}} &
  \multicolumn{1}{c|}{{\textit{I}}}\\
  \hline
2025 Nov 13 & WFI / 1.2$-$m PRL & 14.50 & 1.47 & 2.12 & 3.32 & 24.3 & 13.37 $\pm$ 0.02 & 12.43 $\pm$ 0.01 & -- & --\\
2025 Nov 18 & WFI / 1.2$-$m PRL & 19.50 & 1.57 & 2.05 & 2.88 & 31.5 & 13.29 $\pm$ 0.06 & 12.14 $\pm$ 0.02 & 11.50 $\pm$ 0.01 & --\\
2025 Dec 16 & WFI / 0.5$-$m PRL & 47.50 & 2.22 & 1.80 & 1.17 & 25.7 & 14.46 $\pm$ 0.12 & 13.64 $\pm$ 0.05 & 13.16 $\pm$ 0.09 & 13.05 $\pm$ 0.11 \\
2025 Dec 18 & WFI / 0.5$-$m PRL & 49.50 & 2.28 & 1.80 & 1.15 & 24.6 & 14.52 $\pm$ 0.11 & 13.65 $\pm$ 0.11 & 13.06 $\pm$ 0.05 & --\\
2025 Dec 19 & WFI / 0.5$-$m PRL & 50.50 & 2.31 & 1.80 & 1.08 & 23.8 & 14.66 $\pm$ 0.06 & 13.63 $\pm$ 0.02 & 12.94 $\pm$ 0.05 & 12.89 $\pm$ 0.01\\
2025 Dec 20 & WFI / 0.5$-$m PRL & 51.50 & 2.34 & 1.80 & 1.41 & 23.2 & 14.39 $\pm$ 0.21 & 13.34 $\pm$ 0.08 & 12.77 $\pm$ 0.05 & -- \\
2025 Dec 31 & Atik383L+ / CepO & 63.46 & 2.67 & 1.87 & 1.05 & 15.0 & 14.87 $\pm$ 0.13 & 13.84 $\pm$ 0.06 & -- & --\\
\hline
\end{tabular}
}
\end{threeparttable}
\label{tab: Photometric PRL-I}
\end{table*}

\begin{table*}
\caption{Broad-band Sloan photometric observations obtained with FOC (2.5$-$m) and HFOSC (2.0$-$m). The reported magnitudes are the median values of each dataset.}
\begin{threeparttable}
\setlength{\tabcolsep}{5.5pt}
        \resizebox{\textwidth}{!}{%
\begin{tabular}{ccccccccccc}
\hline
\multicolumn{1}{|c|}{{Date}} &
  \multicolumn{1}{c}{{Instrument/}} &
  \multicolumn{1}{c|}{{TFP}} &
  \multicolumn{1}{c|}{{r$_h$}} &
  \multicolumn{1}{c|}{{$\Delta$}} &
  \multicolumn{1}{c|}{{Airmass}} &
  \multicolumn{1}{c|}{{Phase}} &
  \multicolumn{4}{c|}{\hspace{0.5cm}{Median Magnitudes (mag)}}\\
\multicolumn{1}{c|}{} &
\multicolumn{1}{c|}{Telescope} &
  \multicolumn{1}{c|}{{(Days)}} &
  \multicolumn{1}{c|}{{(au)}} &
  \multicolumn{1}{c|}{{(au)}} &
  \multicolumn{1}{c|}{{}} &
  \multicolumn{1}{c|}{{Angle ($^{\circ}$)}} &
  \multicolumn{1}{c|}{{\textit{u}}} &
  \multicolumn{1}{c|}{{\textit{g}}} &
  \multicolumn{1}{c|}{{\textit{r}}} &
  \multicolumn{1}{c|}{{\textit{i}}}\\
  \hline
2025 Dec 20 & HFOSC / 2.0$-$m HCT & 51.50 & 2.34 & 1.80 & 1.12 & 23.2 & 15.20 $\pm$ 0.22 & 13.76 $\pm$ 0.06 & 13.33 $\pm$ 0.08 & --\\
2025 Dec 29 & FOC / 2.5$-$m PRL & 60.50 & 2.60 & 1.85 & 1.71 & 17.2 & 17.14 $\pm$ 0.01 & 14.79 $\pm$ 0.01 & 14.27 $\pm$ 0.01 & --\\
2026 Jan 06 & FOC / 2.5$-$m PRL & 68.50 & 2.84 & 1.96 & 2.69 & 10.5 & -- & 14.97 $\pm$ 0.02 & 14.37 $\pm$ 0.01 & 14.19 $\pm$ 0.01\\
2026 Jan 16 & FOC / 2.5$-$m PRL & 78.50 & 3.16 & 2.19 & 1.71 & 17.2 & -- & 16.02 $\pm$ 0.01 & 15.50 $\pm$ 0.01 & 15.43 $\pm$ 0.01\\
\hline
\end{tabular}
}

\end{threeparttable}
\label{tab: Photometric PRL-II}
\end{table*}


\subsection{Reflective gradient}\label{sec: Results_4}

The reflective gradient is a parameter that provides insight into the reddening of cometary colours, offering a qualitative estimate of the size of the dust grains in the cometary coma \citep{Ahearn_1984}. It is defined as $(dS/d\lambda)/\Tilde{S}$, where $S$ is the reflectance of the comet, which is the ratio of the comet flux ($F(\lambda)$) to the solar flux ($F_{\odot}(\lambda)$), i.e., $S(\lambda) = \frac{F(\lambda)}{F_{\odot}(\lambda)}$. $d\lambda$ is defined as the difference between two wavelengths, which will be the central bandwidth for different filters, and $\Tilde{S}$ is the mean value of the reflectance in the bandpass \citep{Jewitt_2002}, or it can be written as:

\begin{equation}\label{ref_grad2}
    S'(\lambda) = \frac{2}{\Delta\lambda}\frac{S(\lambda_2)-S(\lambda_1)}{S(\lambda_2)+S(\lambda_1)} 
\end{equation}

where $\Delta\lambda = \lambda_2 - \lambda_1$, is given in \AA. The unit of the reflective gradient, $S'(\lambda)$ is given in $\%$ (1000 \AA)$^{-1}$. In units of magnitude, we get:
\begin{equation}\label{ref_grad3}
    S'(\lambda) = \frac{2}{\Delta\lambda}\frac{10^{0.4(CI_{\rm comet}-CI_{\rm sun})}-1}{10^{0.4(CI_{\rm comet}-CI_{\rm sun})}+1}
\end{equation}

Here, $CI_{\rm comet}$ and $CI_{\rm sun}$ are the comet and solar colours. The solar colours have been taken from \citet{Holmberg_2006}. 

Using equation (\ref{ref_grad3}), the value of the reflective gradient for $B-V$ is found to be $33.36 \pm 10.67$ $\%$ (1000 \AA)$^{-1}$, for $V-R$ to be $20.24 \pm 9.38$ $\%$ (1000 \AA)$^{-1}$, $R-I$ to be $-15.61 \pm 2.60$ $\%$ (1000 \AA)$^{-1}$ and for $B-R$ to be $30.20 \pm 8.64$ $\%$ (1000 \AA)$^{-1}$ . The post-perihelion reflective gradient in the $B-R$ band, which closely matches with $4000-7000 \AA$, is the 1-$\sigma$ error than the pre-perihelion values given by \citet{3I_disc_1,3I_disc_2,Opitom_3I_disc,Belyakov_3I_disc,Fuente_3I,Hoogendam_2025d}.

In the case of Sloan filters, the value of the reflective gradient for $g-r$ is $< 4.38$ $\%$ (1000 \AA)$^{-1}$, and for $r-i$ is $ < 0.33 $ $\%$ (1000 \AA)$^{-1}$. This shows that the comet's reflectivity is redder in the $B-V$, $V-R$, and $g-r$ colours, while it is similar or bluer in the $R-I$ and $r-i$ colours. This suggests a reflectance spectrum that reddens in the optical region and flattens at longer wavelengths.

The comet colours (see Section \ref{sec: Results_3}) and the reflectance gradient are indicative of the cometary coma being rich in ultra-red material normally only found in solar system minor bodies beyond 10 au \citep[see][and references therein]{Jewitt_2019}. However, some care is needed when comparing the colour and reflectance from the dust in a comet's coma with the surface colour of KBOs, as the underlying physical processes are different.
\vspace{-0.4em}
\subsection{Post-perihelion Spin Period of the Comet 3I/ATLAS}\label{sec: Results_5}

\begin{table*}
\caption{Time-Series Photometry Obtained with the Bessel R-Band Filter at CepO. The reported magnitudes are median values of each nightly dataset.}
\begin{threeparttable}
\setlength{\tabcolsep}{3.5pt}
        \resizebox{\textwidth}{!}{%
\begin{tabular}{ccccccccccc}
\hline
\multicolumn{1}{|c|}{{Date}} &
\multicolumn{1}{c|}{{Time}} &
  \multicolumn{1}{c|}{{TFP}} &
  \multicolumn{1}{c|}{{r$_h$}} &
  \multicolumn{1}{c|}{{$\Delta$}} &
  \multicolumn{1}{c|}{{Airmass}} &
  \multicolumn{1}{c|}{{Phase}} &
  \multicolumn{1}{c|}{\hspace{0.5cm}{Median Magnitudes ($R$)}}\\
\multicolumn{1}{|c|}{} &
\multicolumn{1}{c|}{{(UT)}} &
  \multicolumn{1}{c|}{{(Days)}} &
  \multicolumn{1}{c|}{{(au)}} &
  \multicolumn{1}{c|}{{(au)}} &
  \multicolumn{1}{c|}{{}} &
  \multicolumn{1}{c|}{{Angle ($^{\circ}$)}} &
  \multicolumn{1}{c|}{(mag)}\\\hline
2025 Dec 23 & 20:15:36.7 $-$ 23:23:04.9 & 55.41 & 2.42 & 1.81 & 1.11 & 21.1 & 13.01 $\pm$ 0.02\\
2025 Dec 24 & 20:11:32.0 $-$ 23:27:27.5 & 56.41 & 2.45 & 1.81 & 1.09 & 20.4 & 13.05 $\pm$ 0.02\\
2025 Dec 25 & 20:15:37.4 $-$ 23:33:37.9 & 57.42 & 2.48 & 1.82 & 1.07 & 19.6 & 13.10 $\pm$ 0.03\\
2025 Dec 26 & 20:03:18.2 $-$ 23:31:50.5 & 58.41 & 2.52 & 1.82 & 1.07 & 18.9 & 13.14 $\pm$ 0.02\\
2025 Dec 27 & 19:47:20.7 $-$ 22:54:52.5 & 59.39 & 2.54 & 1.83 & 1.09 & 18.1 & 13.19 $\pm$ 0.03\\
2025 Dec 28 & 20:09:52.3 $-$ 23:21:34.1 & 60.41 & 2.57 & 1.84 & 1.05 & 17.3 & 13.24 $\pm$ 0.04\\
2025 Dec 29 & 22:41:10.8 $-$ 00:03:24.0 & 61.48 & 2.61 & 1.85 & 1.07 & 16.5 & 13.26 $\pm$ 0.02\\
\hline
\end{tabular}
}
\end{threeparttable}
\label{tab: Photometric CepO}
\end{table*}

\begin{figure}
	\includegraphics[width=\columnwidth]{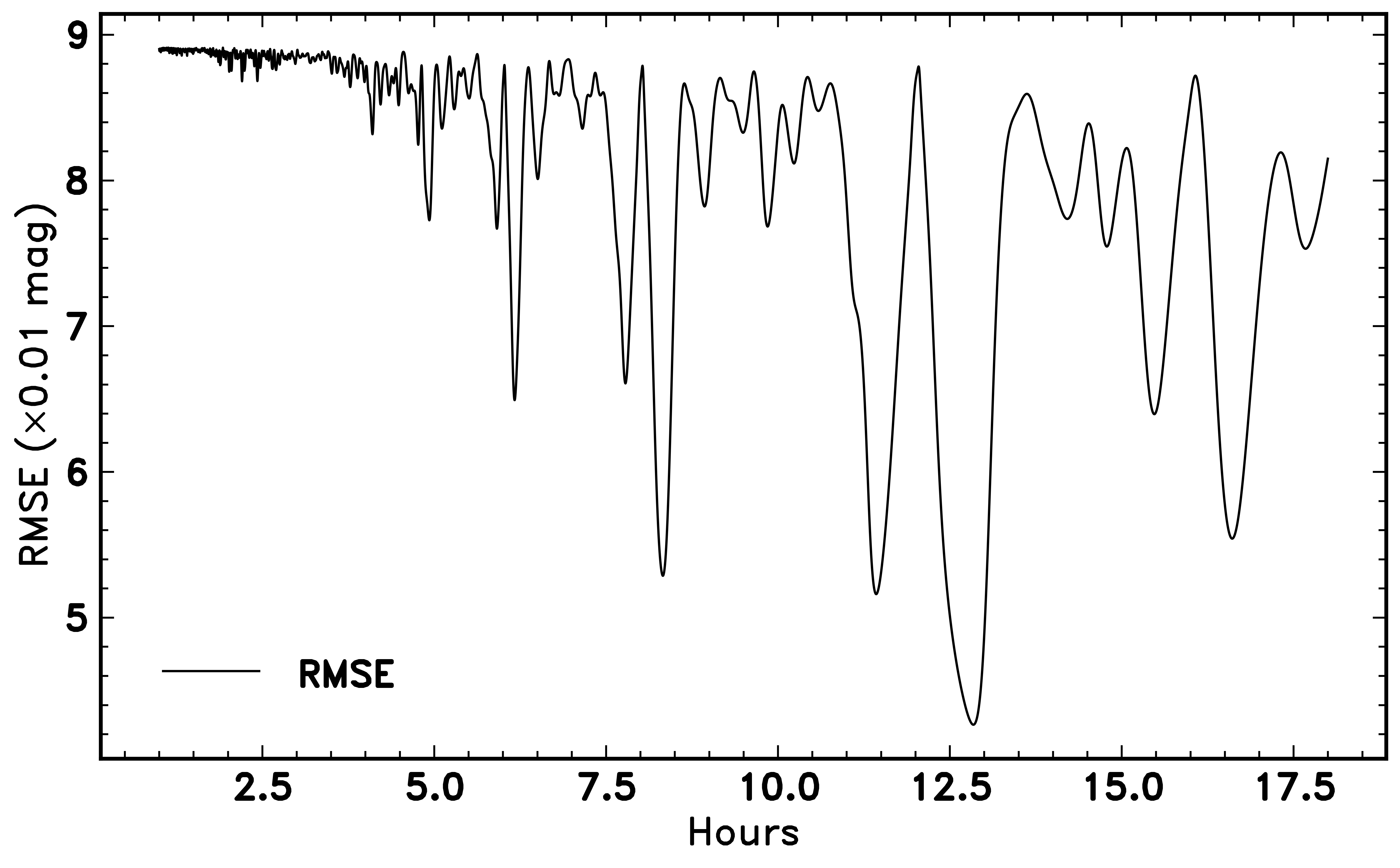}
    \caption{The Root Mean Square Error (RMSE) for the different periods found using the Tycho Software, which is based on Fourier Analysis of Light Curve data from CepO.}
    \label{fig: RMSE}
\end{figure}

The R band photometric data, shown in Table \ref{tab: Photometric CepO}, were analysed using the Tycho light curve software \citep{Parrott_2020,2025epsc.conf.1222P}, which implements the Fourier Analysis of Light Curves (FALC) method \citep{Harris_1989}.  The resulting periodogram (see Fig. \ref{fig: RMSE}) shows a global minimum at 12.81 h. However, there are other prominent minimas, mainly 8.32 h, 11.43 h, and 16.61 h.  The local minima at 16.61 $\pm$ 0.02 h is consistent with previous reports of the spin period by \citet{Tony_2025} and \citet{Fuente_3I}. The CepO data, on their own, cannot be used for estimating the spin period of the comet.  This is mainly due to the sparse filling of the light curve, although spread over many nights of observations.  It is expected that the photometric values will be of use to estimate a spin period once data from other time zones becomes available to fill the gaps.  

\subsection{Nuclear Radius}\label{sec: Results_6}
The nuclear radius of a comet can be estimated by using non-gravitational (hereafter NG) acceleration parameters \citep{Sosa_2011}.

The detailed methodology is described in section 6.5 of \citet{Ahuja_2025_MNRAS}. The method is mainly based on the effect of NG acceleration produced by the sublimation of the volatiles, and it is described as:
\begin{equation}
    \frac{d^2\textbf{r}}{dt^2} = -\frac{GM_\odot \textbf{r}}{r^3} + \nabla \mathcal{R} + A_1 g(r) \hat{\textbf{r}} + A_2 g(r) \hat{\textbf{t}} + A_3 g(r) \hat{\textbf{n}}
\end{equation}
where the first two terms represent the gravitational acceleration, in which $G$ is the gravitational constant, $\mathcal{R}$ is the planetary disturbing function, $M_\odot$ is the Sun's mass, $r$ is the heliocentric distance, and the remaining terms represent the NG accelerations. In this model, the three parameters, i.e., $A_1$, $A_2$, and $A_3$, represent the modulus of the NG acceleration in the radial, transverse, and normal directions at 1 au from the Sun. To account for the asymmetric nature of the coma, there is one additional parameter, $DT$, which represents the time offset of the maximum brightness. $g(r)$ is the empirical sublimation function for the different molecules, given by:
\begin{equation}
    g(r) = \alpha \left(\frac{r}{r_0}\right)^{-m} \left[1 + \left(\frac{r}{r_0}\right)^n \right]^{-k}
\end{equation}
The parameters depend on the dominant sublimating volatile species, i.e., H$_2$O or CO$_2$. Using the details mentioned in \citet{krowlikoska_ngf_2017}, for H$_2$O sublimation, the parametric values are $\alpha = 0.1113$, $m=2.15$, $n=5.093$, $k=4.6142$, $r_0 = 2.808$ au and for CO$_2$ sublimation, the parameters are adopted from the \textsc{NASA JPL} Small-Body Database (\textsc{SBDB}\footnote{\url{https://ssd.jpl.nasa.gov/tools/sbdb_lookup.html\#/?sstr=3I}}) i.e., $\alpha = 1$, $m=2$, $n=0$, $k=0$, and $r_0 = 1$ au. The NG acceleration, $J$ is defined as:
\begin{equation}
    J = \sqrt{A_1^2 + A_2^2 + A_3^2} \times g[r(t-DT)]
\end{equation}
The resulting NG acceleration is then used to estimate the cometary mass.
\begin{equation}
    M_N J = Qmu
\end{equation}
Here, $M_N$ is the mass of the comet, $m$ is the mass of the dominant molecule, $Q(r)$ is the production rate of the molecule, $J$ is the NG acceleration, and the effective gas outflow velocity ($u$) is assumed to be $\frac{0.85}{\sqrt{r_h}}$ km s$^{-1}$ \citep{Cochran_1993}. However, recently \citet{Biver_2026} has reported that the gas expansion velocity of comet 3I is approximately smaller by a factor of two than the value calculated by $\frac {0.85}{\sqrt{r_h}}$ km s$^{-1}$. Therefore, we adopt the revised velocity relation in our calculations.

Here, the two different sublimation models are used to determine the comet's mass. First, using CO$_2$ sublimation. \citet{Cordiner_2025} has obtained the pre-perihelion observations from \textit{JWST} at the heliocentric distance of 3.32 au and geocentric distance of 2.73 au and confirmed that the cometary coma is dominated by CO$_2$. The production rate of CO$_2$ is (9.50 $\pm$ 0.05) $\times$ 10$^{26}$ molecules s$^{-1}$. Similarly, \citet{Belyakov_2026} has observed using \textit{JWST} at post-perihelion for two epochs: 2025 Dec 15 at heliocentric distance of 2.19 au and geocentric distance of 1.80 au, and 2025 Dec 27 at heliocentric distance of 2.54 au and geocentric distance of 1.83 au. The  production rate of CO$_2$ is (8.70 $\pm$ 0.09) $\times$ 10$^{27}$ molecules s$^{-1}$ and (5.42 $\pm$ 0.06) $\times$ 10$^{27}$ molecules s$^{-1}$ respectively.

The NG accelerations at 1 au from the Sun for CO$_2$ sublimation are taken from the \textsc{NASA JPL SBDB} to get the mass of the comet. The values of NG Radial accelerations,  A$_1$ $=$ (5.32 $\pm$ 0.12) $\times$ 10$^{-8}$ au day$^{-2}$, Transverse, A$_2$ $=$ (1.15 $\pm$ 0.21) $\times$ 10$^{-8}$ au day$^{-2}$, Normal, A3 $=$ ($-6.85 \pm 0.18$) $\times$ 10$^{-9}$ au Day$^{-2}$, DT $=$ 9.48 days. This gives the mass of the comet to be ($4.85 \pm 1.06$) $\times$ 10$^{11}$ kg. After that, using the equation given below, it can be further used to calculate the radius of the comet. 
\begin{equation}
    R_N = \left(\frac{3M_N}{4\pi \rho}\right)^{1/3}
\end{equation}
Assuming a bulk density as seen for solar system comets \citep{rosetta_2018} as $\rho$ = $537.8 \pm 0.6$  kg m$^{-3}$, the radius of the comet 3I is estimated to be 0.60 $\pm$ 0.04 km. However, its true density may differ owing to its extrasolar origin.

For completeness, we have also used the H$_2$O sublimation model to calculate the radius. The asymmetric NG acceleration parameters corresponding to the H$_2$O sublimation are taken from \citet{Ahuja_2026}. The H$_2$O production rates are adopted from \citet{Cordiner_2025, Belyakov_2026, Tan_2026}, and also from \citet[][references therein]{Biver_2026}. We have calculated H$_2$O production rates from the TRAPPIST OH filter provided by \citet{Jehin_17515_2025,Jehin_17538_2025}, using the relation Q(H$_2$O) = 1.1 $\times$ Q(OH) \citep{Crovisier_1989}. Following similar calculations as for CO$_2$ sublimation, the mass of the comet is found to be ($3.66 \pm 0.53$) $\times$ 10$^{11}$ kg, and the nuclear radius is found to be 0.55 $\pm$ 0.10 km. 

Hence, considering the 1$\sigma$ uncertainties from the different methods, the nuclear radius of the comet is constrained to lie between 0.45 and 0.65 km. This range is consistent with the values reported by \citet{Forbes_2026} in the literature (see Table \ref{tab:nuclear_radius} and Table 3 of \citet{Hui_2026}).

\begin{table}
\centering
\caption{Estimated nuclear radius of comet 3I obtained using the NG method and the different published results.}
\label{tab:nuclear_radius}
\setlength{\tabcolsep}{3pt}
\resizebox{\linewidth}{!}{%
\begin{tabular}{lcc}
\hline
Method & Radius (km) & Reference / Data Source \\
\hline
NG force model (CO$_2$ sublimation) & $0.60 \pm 0.04$ & This work \\
NG force model (H$_2$O sublimation) & $0.55 \pm 0.10$ & This Work \\
\hline
Adopted range (1$\sigma$) & $0.45 - 0.65$ & This work \\
\hline
Surface brightness distribution & $0.22 - 2.8$ & \citet{Jewitt_2025} \\
NG force model & $0.41 - 0.53$ &  \citet{Forbes_2026} \\
Nucleus Extraction & $1.3 \pm 0.2$ &  \citet{Hui_2026} \\
NG force model & $1.5 \pm 0.1$ &  \citet{Hui_2026} \\
\hline
\end{tabular}
}
\end{table}

\section{Conclusions}\label{sec: Conclusion}
Post-perihelion observations of the interstellar comet 3I provide insight into its physical and chemical properties. Using the observations, we conclude the following:

\begin{itemize}
    \item The post-perihelion spectroscopic observations of the comet 3I confirm the change in the chemical composition of the comet, which was seen to be carbon-depleted at pre-perihelion, is now identified as a carbon-typical comet, as both C$_2$/CN and C$_3$/CN have changed from their pre-perihelion values. This change is likely due to exposure of the comet's surface layers to Galactic Cosmic Rays (GCR), which could explain the depleted composition observed in the pre-perihelion values.
    \item The comet has been found to be dust-rich based on its $Af\rho/Q(\mathrm{CN})$ values.
    \item The post-perihelion broad-band and narrow-band colours closely match the pre-perihelion colours, indicating that the change in the production rate and even its composition does not change the optical colours. This is confirmed by comparing the photometric magnitudes calculated from the comet spectrum in the presence of molecular emissions with those from a pure continuum. 
    \item The reflectivity of the comet is found to be very red in the case of $B-V$, $V-R$, and $g-r$ colours, while it is similar or bluer in the $R-I$ and $r-i$ colours as compared to solar values. Hence, the reflectance spectrum is reddened in the optical region and flattens at longer wavelengths. It is indicative of the comet being rich in ultra-red material, as seen in a few of the objects in the solar system.
    \item R-band time series photometry reported here can help fill gaps in the time coverage and constrain the spin period. 
    \item Using NG accelerations, the nuclear radius of the comet is constrained to be in the range of 0.45 $-$ 0.65 km.
\end{itemize}

\section*{Acknowledgements}

We thank the anonymous reviewer for their useful suggestions, which have significantly improved the manuscript. Work at the Physical Research Laboratory is supported by the Department of Space, Govt. of India. GA was a senior research fellow, PS and B. Ailawadhi were Post-doctoral fellows, and B. Arvind was a Junior Research Fellow at PRL during the time of this work. We acknowledge the local staff at the Mount Abu Observatory for their support.

We would like to thank the Director, IIA,  for allowing us to use the Director's Discretionary Time on the HCT for part of the observations presented in this work. We thank the staff of IAO, Hanle and CREST, Hosakote, who made these observations possible. The facilities at IAO and CREST are operated by the Indian Institute of Astrophysics, Bangalore. 

This work is a result of the bilateral Belgo-Indian projects on Precision Astronomical Spectroscopy for Stellar and solar system bodies, BIPASS, funded by the Belgian Federal Science Policy Office (BELSPO, Government of Belgium; BL$/$33$/$IN22\texttt{\_}BIPASS) and the International Division, Department of Science and Technology, (DST, Government of India; DST/INT/BELG/P-01/2021(G)).

The authors thank Daniel Parrott for the discussion on the technical details of the Tycho software. The authors also acknowledge the use of ChatGPT (OpenAI) for helping improve the language and refining plotting codes during the preparation of this manuscript.

\section*{Data Availability}
The data underlying this article will be shared on reasonable request to the corresponding author.



\bibliographystyle{mnras}
\bibliography{reference} 




\appendix
\clearpage
\onecolumn 
\section{\texorpdfstring{Variation of gas production rates with heliocentric distance and days after perihelion and the power law fitting with heliocentric distance in comet 3I/ATLAS}{Variation of gas production rates with heliocentric distance and days after perihelion and the power law fitting with heliocentric distance in comet 3I/ATLAS}}\label{Qs}
\begin{table}
\centering
\caption{Power-law fits of the form $Q = A\,r^{k}_{\mathrm{h}}$ to the production rates of CN, C$_2$, and C$_3$ based on our observations and on the combined datasets \citep{Jehin_17515_2025,Jehin_17538_2025,Hoogendam_2026,Zhao_2026} as a function of heliocentric distance in the post-perihelion phase.}
\label{tab: Power Law}
\setlength{\tabcolsep}{15pt}
\resizebox{\linewidth}{!}{%
\begin{tabular}{lcccc}
\hline
 & \multicolumn{2}{c}{Our dataset} & \multicolumn{2}{c}{Combined dataset} \\
\cline{2-5}
Species & $A$ & $k$ & $A$ & $k$ \\
\hline
CN  & $(4.41 \pm 0.92)\times10^{26}$ & $-3.72 \pm 0.37$ & $(8.04 \pm 1.71)\times10^{26}$ & $-5.07 \pm 0.32$ \\
C$_2$ & $(2.58 \pm 0.55)\times10^{26}$ & $-3.25 \pm 0.45$ & $(3.21 \pm 0.71)\times10^{26}$ & $-3.84 \pm 0.35$ \\
C$_3$ & $(3.75 \pm 1.24)\times10^{25}$ & $-3.14 \pm 0.62$ & $(0.78 \pm 0.24)\times10^{25}$ & $-1.67 \pm 0.42$ \\
\hline
\end{tabular}
}
\end{table}

\begin{figure*}
\begin{subfigure}{0.49\linewidth}
\centering
\includegraphics[width = 0.95\textwidth]{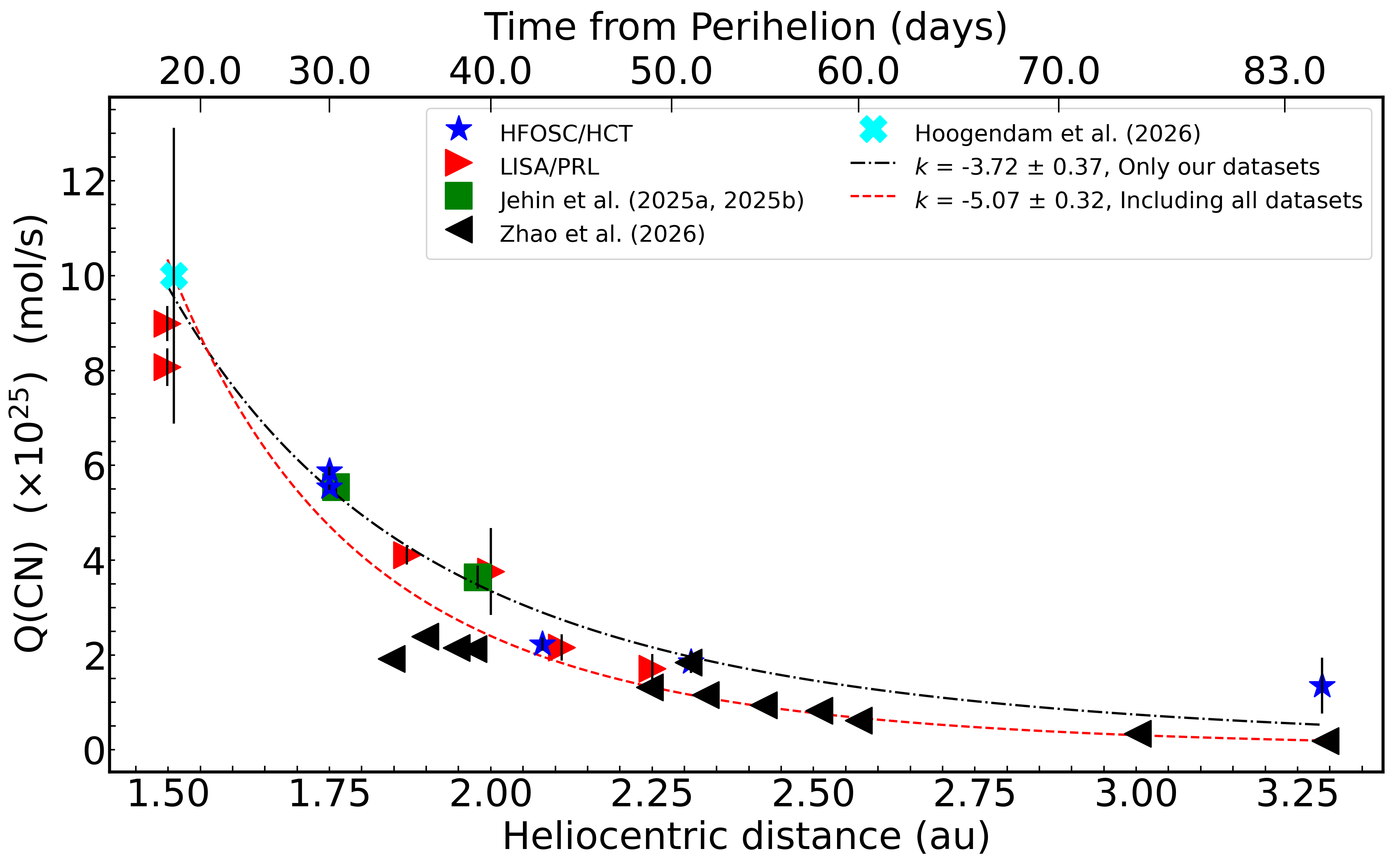}
\subcaption{Power law fitting on the activity trend of Q(CN).}
\label{QCN_fit}
\end{subfigure}\hfill
\begin{subfigure}{0.49\linewidth}
\centering
\includegraphics[width = 0.95\textwidth]{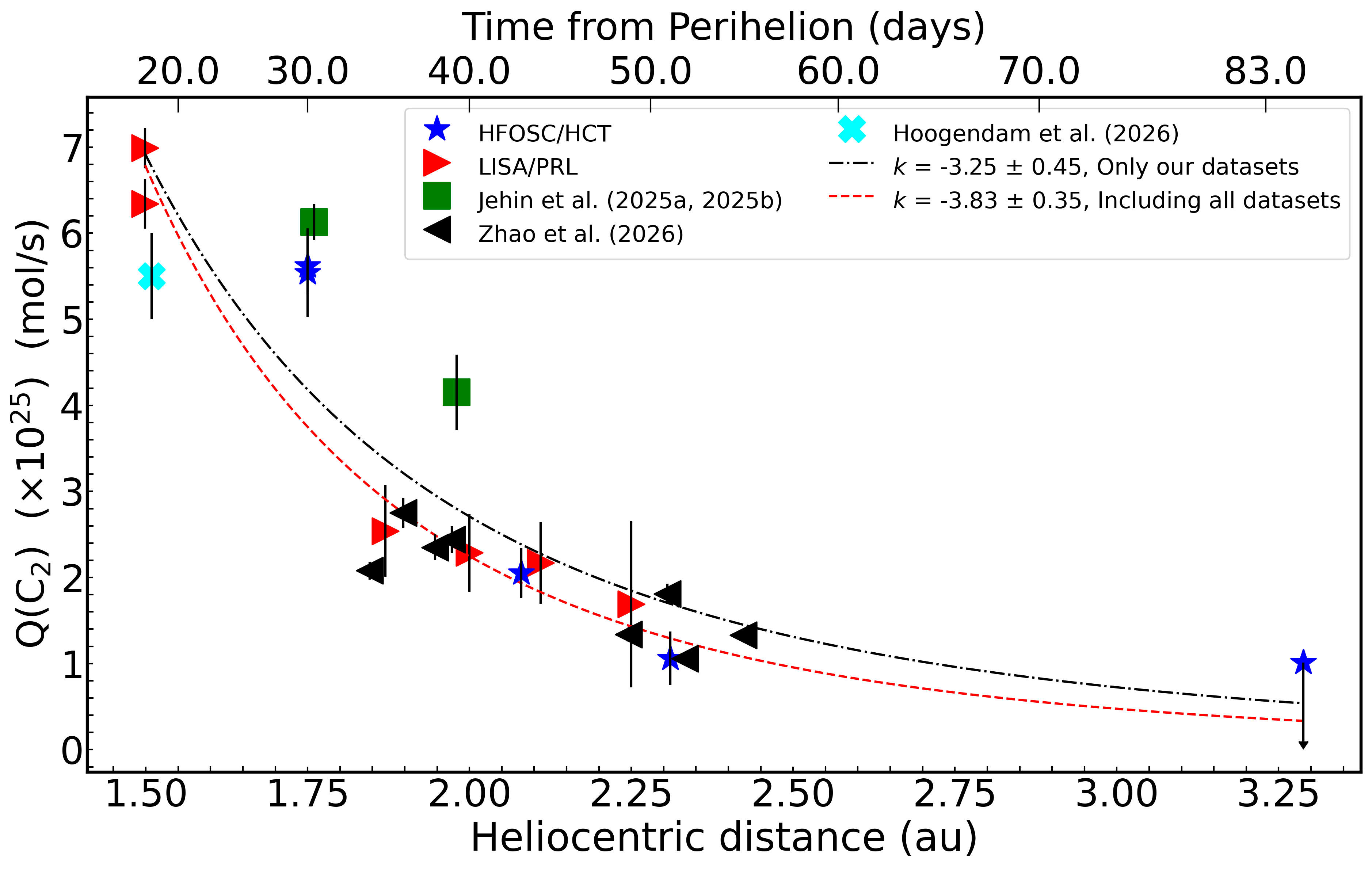}
\subcaption{Power law fitting on the activity trend of Q(C$_2$).}
\label{QC2_fit}
\end{subfigure}\hfill
\begin{subfigure}{0.49\linewidth}
\centering
\includegraphics[width = 0.95\textwidth]{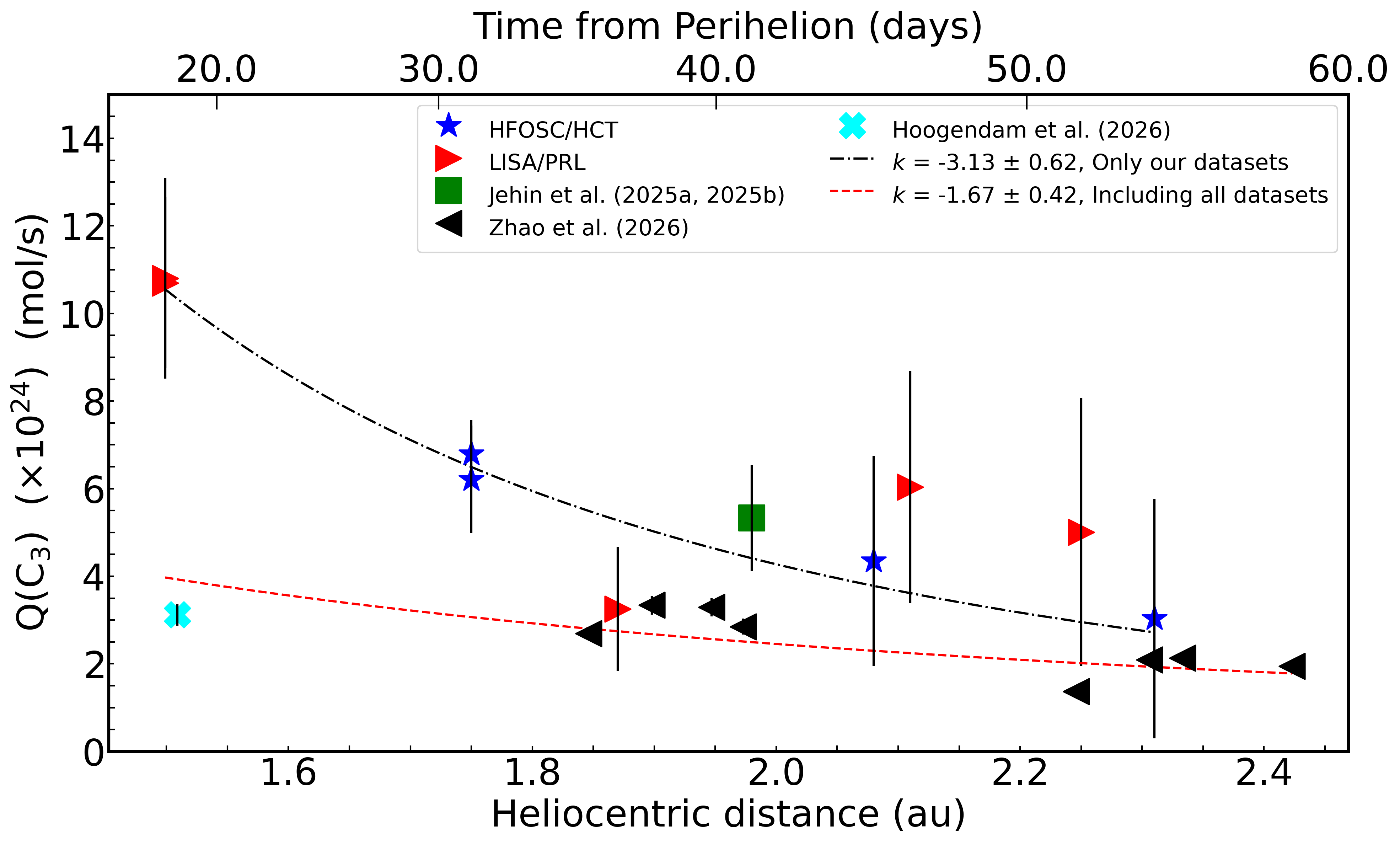}
\subcaption{Power law fitting on the activity trend of Q(C$_3$).}
\label{QC3_fit}
\end{subfigure}\hfill
\caption{Activity trends and the respective power law fitting of comet 3I/ATLAS as a function of heliocentric distance (au) and t. The observation from LISA/PRL is marked as \textit{red right triangle}, from HFOSC/HCT is marked as \textit{blue asterisk}, from \citet{Jehin_17515_2025,Jehin_17538_2025} marked as \textit{green square}, from \citet{Zhao_2026} marked as \textit{black left triangle}, and the observation from \citet{Hoogendam_2026} marked as \textit{cyan cross}.}
\end{figure*}  

\clearpage
\section{\texorpdfstring{A\textit{\MakeLowercase{f}}$\rho$ measurements and LS-processed image of comet 3I/ATLAS}{Afrho measurements and LS-processed image of comet 3I/ATLAS}}\label{dust}

\begin{figure*}
\begin{subfigure}{0.49\linewidth}
\centering
\includegraphics[width = 0.95\textwidth]{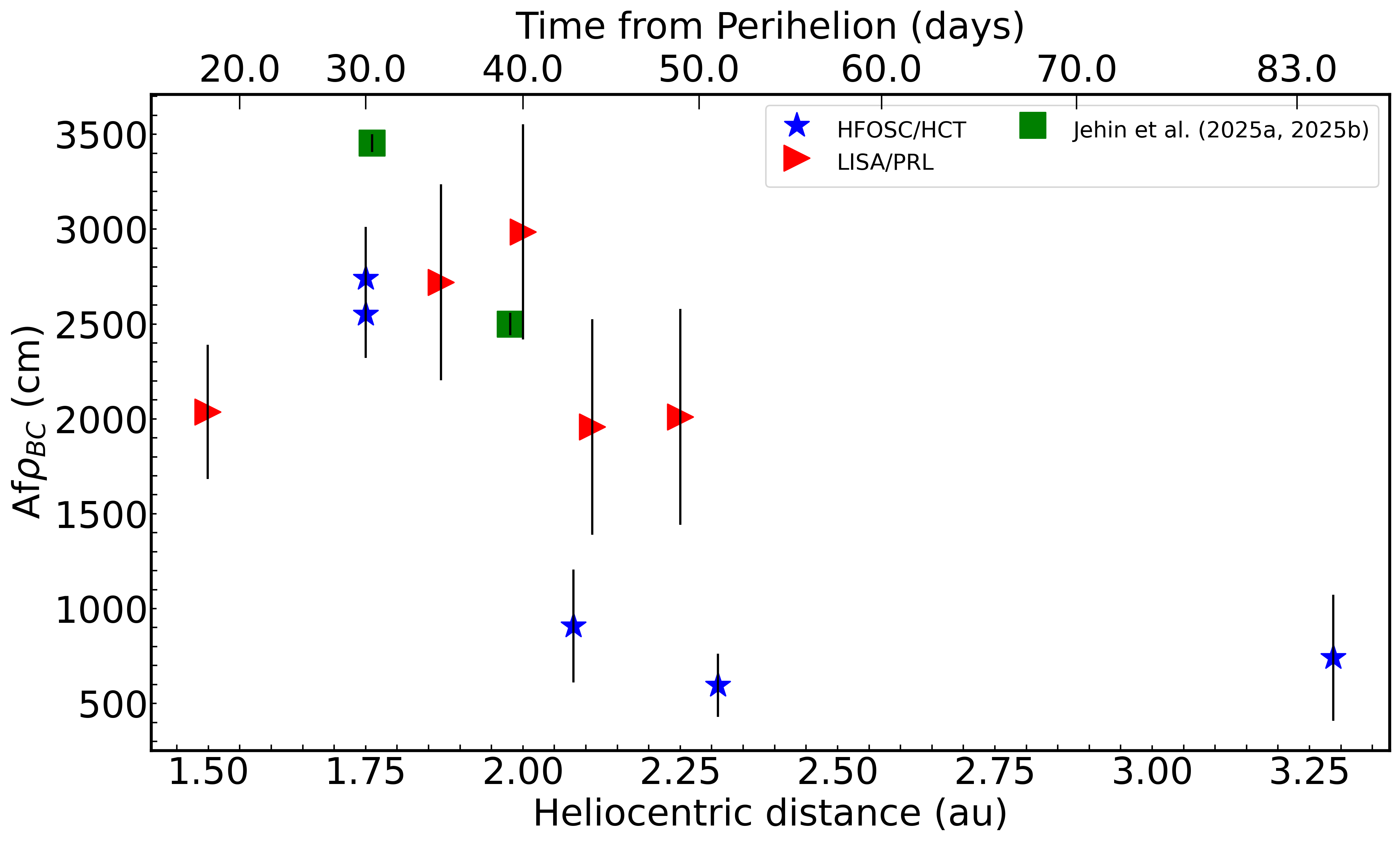}
\subcaption{Activity trend of A$(\theta = 0^{\circ})f\rho$ (BC).}
\label{A_BC}
\end{subfigure}\hfill
\begin{subfigure}{0.49\linewidth}
\centering
\includegraphics[width = 0.95\textwidth]{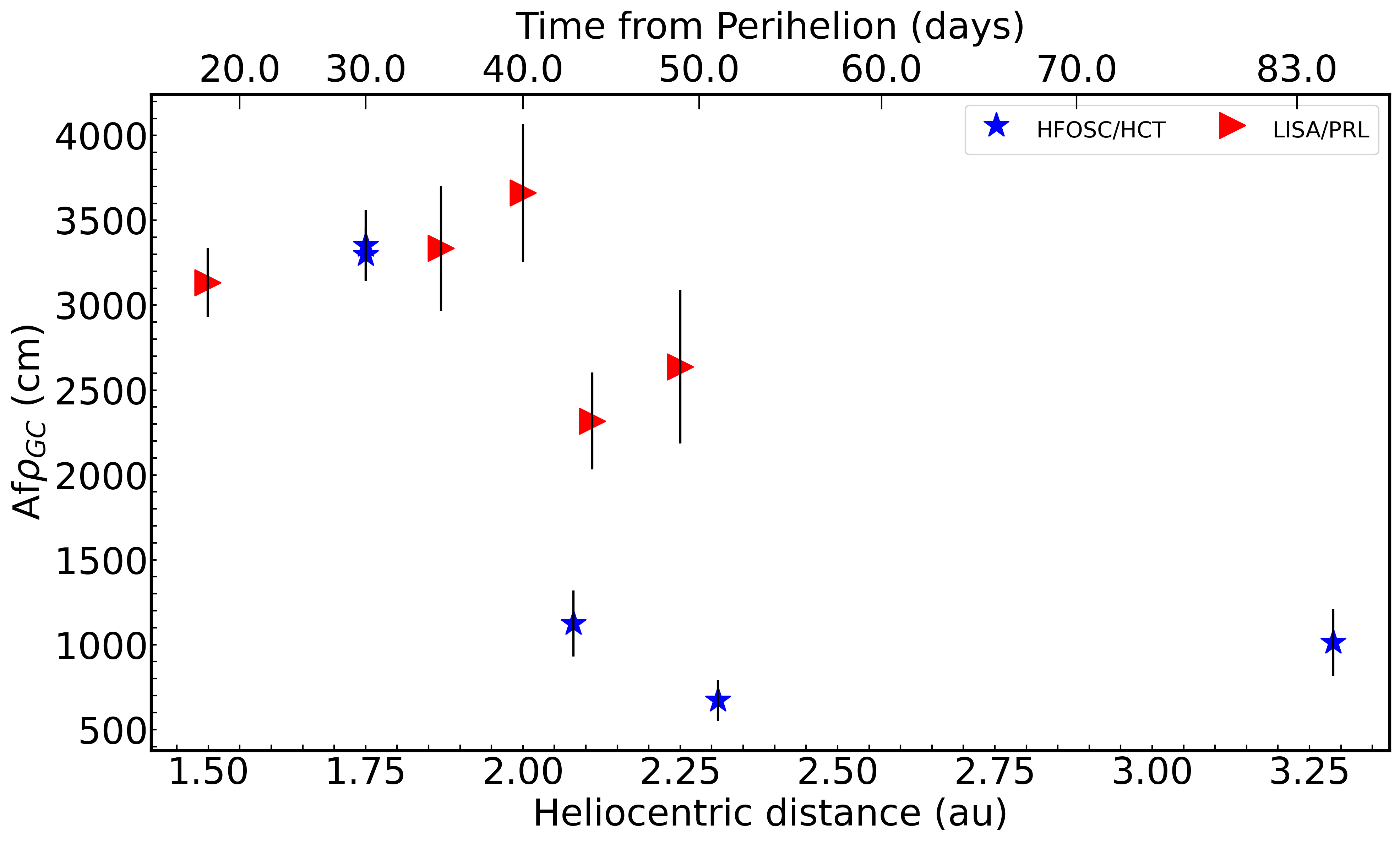}
\subcaption{Activity trend of A$(\theta = 0^{\circ})f\rho$ (GC).}
\label{A_GC}
\end{subfigure}
\caption{Activity trends of comet 3I/ATLAS as a function of days to perihelion and heliocentric distance (au). The observation from LISA/PRL is marked as red right triangle, from HFOSC/HCT is marked as blue asterisk, from \citet{Jehin_17515_2025,Jehin_17538_2025} marked as green square.}
\end{figure*}

\begin{figure}
    \centering
    \includegraphics[width=0.6\linewidth]{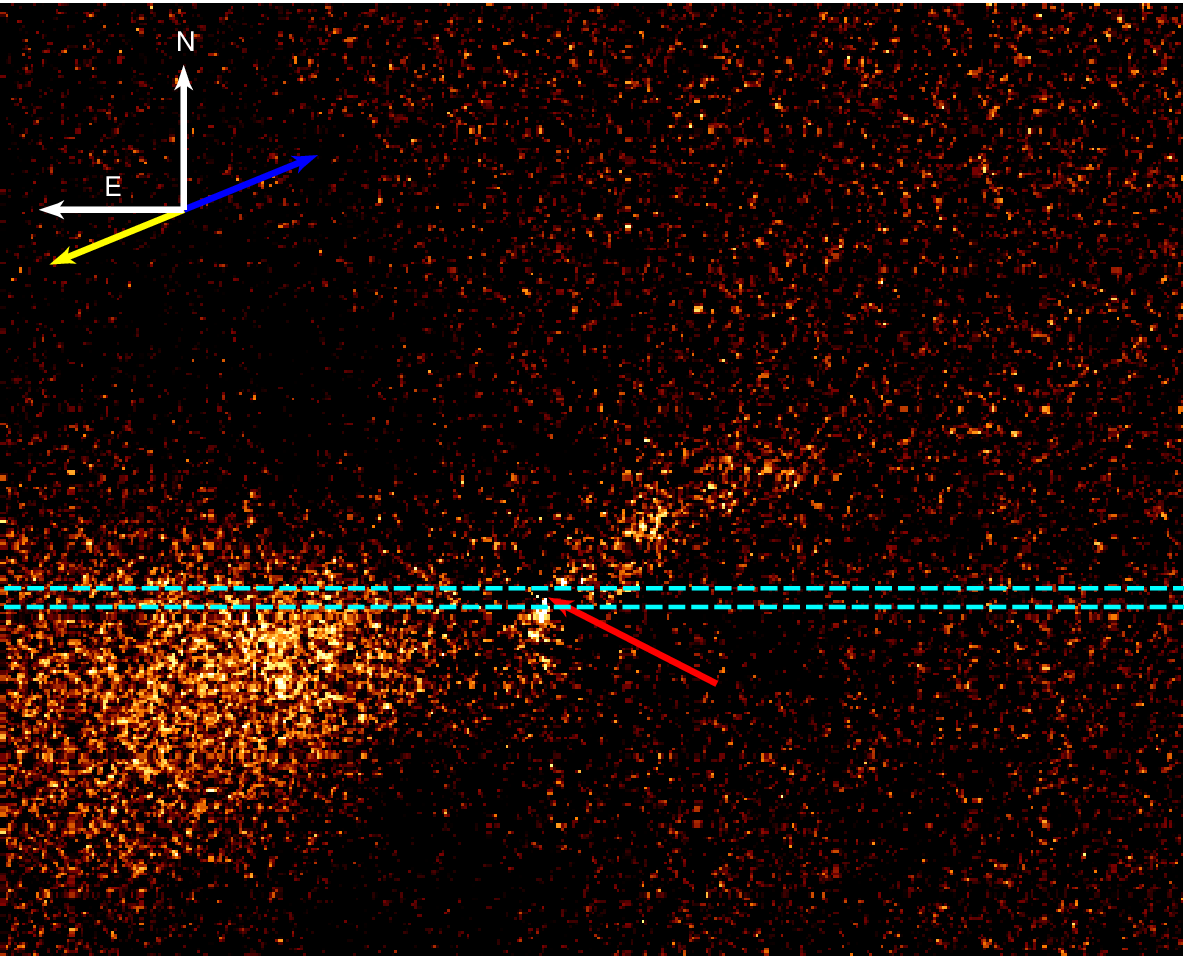}
    \caption{Larson--Sekanina (LS) processed image obtained on 2025 Dec 20 at $45^\circ$ rotation. The slit orientation is marked by a dashed cyan line. The spectrograph slit is oriented along the east--west direction. The red arrow marks the photocentre. The dust-tail direction and the Sun's direction are indicated by blue and yellow arrows, respectively.}
    \label{fig: LS_afp}
\end{figure}


\bsp	
\label{lastpage}
\end{document}